\documentclass{aa}\usepackage{psfig}
\include{epsf}

\def\rmd{{\rm d}}
\def\mpcoh{\,h^{-1}\,{\rm Mpc}}
\def\kpcoh{\,h^{-1}\,{\rm kpc}}
\newcommand{\be}{\begin{equation}}
\newcommand{\ee}{\end{equation}}
\newcommand{\ba}{\begin{eqnarray}}
\newcommand{\ea}{\end{eqnarray}}

\newcommand{\kb}{\mbox{\boldmath $\kappa$}}
\newcommand{\bone}{\mbox{\boldmath $1$}}
\newcommand{\mub}{\mbox{\boldmath $\mu$}}

\newcommand{\bG}{\mbox{\boldmath $G$}}

\newcommand{\nn}{\nonumber \\}
\def\bib{\parskip=0pt\par\noindent\hangindent\parindent
    \parskip =2ex plus .5ex minus .1ex}

\begin{document}

\title{Lens magnification by CL0024+1654 in the $U$ and $R$ band}
\titlerunning{Lens magnification by CL0024+1654 in the $U$ and $R$ band}
\authorrunning{Dye et al.}

\author {S. Dye\inst{1},  
	A.N. Taylor\inst{2}, 
	T.R. Greve\inst{3}$^,$\inst{4},
	\"{O}.E. R\"{o}gnvaldsson\inst{5},
	E. van Kampen\inst{2},
	P. Jakobsson\inst{6},
	V.S. Sigmundsson\inst{6},
	E.H. Gudmundsson\inst{6},
	J. Hjorth\inst{3}}

\institute{
\inst{1} Astrophysics Group, Blackett Laboratory, Imperial College,
	Prince Consort Road, London SW7 2BW, U.K. \\
\inst{2} Institute for Astronomy, University of Edinburgh,
	Royal Observatory, Blackford Hill, Edinburgh EH9 3HJ, U.K.\\
\inst{3} Astronomical Observatory, Juliane Maries Vej 30, 
	DK-2100 Copenhagen \O, Denmark\\
\inst{4} Department of Physics \& Astronomy, University College London,
	Gower Street, London WC1E 6BT, UK \\
\inst{5} NORDITA, Blegdamsvej 17, DK-2100 Copenhagen \O, Denmark\\
\inst{6} Science Institute, Dunhaga 3, IS-107 Reykjavik, Iceland }

\offprints{Simon Dye, \email{sdye01@ic.ac.uk}}

\date{Received Month XX, 2002 / Accepted Month XX, 2002}

\abstract{ We estimate the total mass distribution of the galaxy
cluster CL0024$+$1654 from the measured source depletion due to lens
magnification in the $R$ band.  Within a radius of $0.54\mpcoh$, a
total projected mass of $(8.1\pm 3.2)\times 10^{14} h^{-1}{\rm
M}_{\odot}$ (EdS) is measured. The $1\sigma$ error here includes shot
noise, source clustering, uncertainty in background count
normalisation and contamination from cluster and foreground
galaxies. This corresponds to a mass-to-light ratio of
M/L$_B=470\pm180$. We compute the luminosity function of CL0024$+$1654
in order to estimate contamination of the background source counts
from cluster galaxies.  Three different magnification-based
reconstruction methods are employed: 1) An estimator method using a
local calculation of lens shear; 2) A non-local, self-consistent
method applicable to axi-symmetric mass distributions; 3) A non-local,
self-consistent method for derivation of 2D mass maps. We have
modified the standard single power-law slope number count theory to
incorporate a break and applied this to our observations.  Fitting
analytical magnification profiles of different cluster models to the
observed number counts, we find that CL0024$+$1654 is best described
either by a NFW model with scale radius $r_s=334\pm191\kpcoh$ and
normalisation $\kappa_s=0.23\pm0.08$ or a power-law profile with slope
$\xi=0.61\pm0.11$, central surface mass density $\kappa_0=1.52\pm0.20$
and assuming a core radius of $r_{core}=35\kpcoh$. The NFW model
predicts that the cumulative projected mass contained within a radius
$R$ scales as ${\rm M}(<R)=2.9\times 10^{14}(R/1')^{1.3-0.5\lg
(R/1')}h^{-1}{\rm M}_{\odot}$. Finally, we have exploited the fact
that flux magnification effectively enables us to probe deeper than
the physical limiting magnitude of our observations in searching for a
change of slope in the $U$ band number counts. We rule out both a total
flattening of the counts with a break up to $U_{\rm AB}\leq 26.6$ and
a change of slope, reported by some studies, from $\rmd \log N /
\rmd m=0.4 \rightarrow 0.15$ up to $U_{\rm AB}\leq26.4$ with 95\%
confidence.
}

\maketitle

\section{Introduction}

The lensing cluster CL0024$+$1654 ranks as one of the most highly
studied clusters to date. Lying at a redshift of $z=0.39$, early
measurements of the cluster's velocity dispersion of $\sigma \simeq
1300 \pm 100 {\rm km \,s^{-1}}$ (Dressler, Schneider \& Gunn 1985)
suggested a formidable mass.  The discovery of a large
gravitationally lensed arc from a blue background galaxy by Koo (1988)
has since provoked a range of studies to estimate the cluster's mass
based on its lensing properties.

The first lens inversion of CL0024$+$1654, by Kassiola, Kovner \& Fort
(1992), noted a violation of the `length theorem' (Kovner 1990) that
the length of the middle segment of the arc should equal the sum of
the other two. The authors demonstrated that a concentration of large
cluster galaxies near the arc centre is necessary to cause this by
perturbing the cluster cusp and were thus able to constrain the
potential of the cluster and the perturbing galaxies. A later analysis
by Bonnet, Mellier \& Fort (1994) constrained the cluster's mass
profile more tightly with the first measurement of weak shear out to a
radius of $1.5\mpcoh$. Wallington, Kochanek \& Koo (1995) confirmed
the perturbing galaxy hypothesis of Kassiola, Kovner \& Fort by
fitting a smooth elliptical cluster potential with two superimposed
$L_*$ galaxy potentials near the centre of the arc.

By parameterising the source and lens models and fitting to six images
of the lensed source galaxy, Kochanski, Dell' Antonio \& Tyson (1996)
again showed that the mass profile of CL0024$+$1654 is consistent with
a smooth isothermal distribution. Furthermore, they found that the
cluster mass traces light fairly well out to a radius of $0.5\mpcoh$
and were able to rule out the existence of any significant
substructure larger than $15\kpcoh$ in the central region.  Using HST
images of the cluster, Tyson, Kochanski \& Dell' Antonio (1998, TKD
hereafter) isolated eight well-resolved images of the blue background
galaxy to construct a high resolution mass map of the cluster.  Their
parametric inversion concluded that more than 98\% of the mass
concentration excluding that contributed from discrete galaxies was
represented by a smooth distribution of mass with a shallower profile
than isothermal.

The most recent lensing analysis of CL0024$+$1654 to date is that by
Broadhurst et al. (2000, B00 hereafter) who provide the first
measurement of the redshift of the blue background galaxy at
$z=1.675$. This redshift breaks the mass-redshift degeneracy present
in all mass estimates of the cluster thus far. Their fit of NFW
profiles (Navarro, Frenk \& White, 1997) to the eight brightest
cluster members is found to be an adequate model to explain the
positions of the five main lensed images. This, they suggest,
highlights the possibility that sub-structure has not been erased in
the cluster.

The first of only two measurements of the cluster's lens magnification
to currently exist was that investigated by Fort, Mellier \&
Dantel-Fort (1997). This was the first detection of depletion in
background galaxy number counts due to geometrical magnification as
predicted by Broadhurst, Taylor \& Peacock (1995, BTP
hereafter). Rather than reconstruct cluster mass, this work
concentrated on characterising the radial distribution of critical
lines to infer the redshift range of the background populations in $B$
and $I$. Using this data, van Kampen (1998) produced an estimate of
the mass of CL0024$+$1654 from lens magnification. 

The only other magnification analysis of the cluster published so far
is that of R\"{o}gnvaldsson et al. (2001; R01 hereafter) using $R$ and
$U$ band observations. Their choice to observe in $U$ was inspired by
the findings of Williams et al. (1996) which suggested a flattening in
the $U$ band number count slope at faint magnitudes. Such a break is
reported to occur at $U_{\rm AB}\simeq 25.5 - 26$ with a change of
slope of $\rmd \log N/\rmd m \simeq 0.4 \rightarrow 0.15$. Given
suitably deep $U$ band imaging, this should therefore manifest itself
as a depletion in the number density of galaxies observed in the
presence of lens magnification (see Section
\ref{sec_dual_slope_mag}). R01 claimed to have detected depletion in
$U$ implying that a break in the slope must be present.

Further investigation has since highlighted concern regarding the
reliability of faint objects extracted by R01 from the $R$ and $U$
band observations. While this is not a large concern for $R$ where the
depletion signal is strong, the claim of detection of depletion in $U$
and hence the reported break in the number count slope is strongly
affected. A re-examination of these findings is therefore necessary.

The motivation driving the paper presented here is severalfold.
Firstly, there is the need to re-evaluate the existing results of
R01. We take the observations from R01 but create new, more reliable
$R$ and $U$ band object catalogues. Using these, we re-calculate the
depletion signal in both bands and use this to re-fit the isothermal
lens model of R01.  Secondly, we wish to extend the analysis of R01:
1) We fit a power-law profile and a NFW model to the depletion
profiles.  2) We transform the measured depletion into mass estimates
using three recently developed magnification reconstruction methods;
the local estimator method of van Kampen (1998) applied by Taylor et
al. (1998, T98 hereafter), the non-local axi-symmetric solution of T98
for reconstruction of radial mass profiles and the non-local method of
Dye \& Taylor (1998) for determining 2D mass maps. 3) We investigate
the relationship between mass, light and galaxy number density in the
cluster. 4) We quantify contamination of our background source sample
by cluster and foreground objects and incorporate this into our
analyses.  5) Finally, we apply two methods to search for a change in
slope in the $U$ band field galaxy number counts by exploiting the
fact that lens magnification enables us to effectively see deeper than
the physical magnitude limit imposed by the observations.

The following section briefly describes data acquisition, reduction,
object extraction and mask generation. Contamination of the source
counts from cluster and foreground galaxies is estimated.  Section
\ref{sec_recon_theory} details the reconstruction methods used in this
paper. In Section \ref{sec_no_count_fitting}, we consider the
properties of the $R$ band background galaxy population and fit
cluster mass models to the measured number count profile. This is
necessary for determination of the mass results presented in Section
\ref{sec_results}. In Section \ref{sec_U_band_pop}, we investigate the
$U$ band galaxy population and test for the existence of a break in
the number counts. Finally, we discuss and summarise our findings in
Section \ref{sec_summary}.

\section{The data}
\label{sec_the_data}

Data acquisition and reduction are described in full detail in R01.  
Here, we highlight the key deviations in the generation of the object
catalogues of this paper from those in R01:
\begin{itemize}

\item A Gaussian smoothing kernel with a FWHM equal to the seeing was
used to extract objects (see Section \ref{sec_obj_extr}). This
compares with the narrower smoothing kernel of FWHM equal to slightly
larger than half the seeing used by R01 (note the incorrect statement
that their smoothing kernel is equal to the seeing).  Matching the width of the
smoothing kernel to the seeing allows optimal extraction in the sense
that spurious detections due to background noise are kept as small as
possible while maintaining a sufficiently large detection success rate.
This creates the most noticeable difference between this
dataset and that of R01; our $U$ and $R$ catalogues contain fewer
faint objects (our total paired object catalogue contains 
29\% fewer objects).

\item Galactic extinction corrections have been applied to both the
$U$ and $R$ band, brightening all objects (see Section
\ref{sec_aq_and_compl}).

\item A more thorough determination of the background noise has been
obtained. The mean RMS background variation has been calculated within
a circular aperture of diameter equal to the seeing FWHM.  The result
of this is that the `$3\sigma$' detection limits calculated in R01
from Poission statistics are fainter than those in this paper by $1.3$
mag in $U$ and $1.7$ mag in $R$ including the galactic extinction
correction.

\item A linear co-ordinate transformation has been
applied to the $U$ band object positions resulting in a more accurate
mapping to the $R$ band. This has improved the position coincidence
matching between both bands.


\end{itemize}

\subsection{Acquisition and completeness}
\label{sec_aq_and_compl}

CL0024$+$1624 was observed in the Cousins $U$ and Cousins $R$ bands
with a $6' \times 6'$ field of view using the Nordic Optical Telescope
in August 1999.  A total integration time of $37$ ksec and seeing
$1.1''$ was obtained in the $U$ band compared with $8.7$ ksec and
$1.0''$ in the $R$. Following R01, we use AB magnitudes throughout
this paper converting from the Cousins $U$ and $R$ magnitudes with an
offset of $+0.71$ mag and $+0.20$ mag respectively (Fukugita et
al. 1995). In addition, we have corrected the data for galactic
extinction amounting to $-0.15$ mag and $-0.31$ mag in the $R$ and $U$
band respectively (Schlegel et al. 1998).  The limiting magnitudes
corresponding to a signal-to-noise ratio of 3 within a seeing disk are
$U_{\rm AB}=25.7$ mag and $R_{\rm AB}=25.8$ mag, with the noise level
taken as the mean RMS of the background. All photometry was performed
using SExtractor (Bertin \& Arnouts 1996; see Section
\ref{sec_obj_extr}).  Magnitudes were measured as either total or
corrected isophotal depending on the proximity of neighbouring
objects. This is managed automatically via SExtractor's {\tt
MAG\_BEST} magnitude definition.

Completeness was estimated from the detectability of synthetic objects
of varying brightness added to the images. Further details can be
found in R01. In re-applying this process to the data of this paper,
we find that the completeness at the $3\sigma$ detection limit is
$84\%$ at $U_{\rm AB}=25.7$ mag and $81\%$ at $R_{\rm AB}=25.8$ mag.

\subsection{Object extraction}
\label{sec_obj_extr}

Objects were extracted from the final reduced images using SExtractor.
With a detection threshold of $1\sigma$ above background and a
Gaussian filtering kernel of FWHM equal to the seeing, catalogues of
all objects with at least 10 connecting pixels brighter than the
threshold were generated. A total of 1887 objects in the $R$ band and
1122 objects in the $U$ band were detected. These totals include stars
but exclude objects classified by SExtractor as having saturated
pixels, being truncated or possessing corrupt isophotal data.  By
matching the 30 brightest stars between both bands, a linear
co-ordinate transformation was calculated and applied to map the $U$
band catalogue onto the $R$ band. Objects were paired within a
positional tolerance equal to the seeing and yielded a total of 875
objects excluding stars. Those objects with star/galaxy classification
indices larger than 0.95 were assumed to be stars and excluded from
the analysis below.

\subsection{Object selection}
\label{sec_obsel}

Separation of the background galaxies from the foreground and cluster
galaxies must be achieved before lens magnification can be evaluated.
Segregation of the cluster and foreground objects is important to
enable estimation of their background sky obscuration which affects
binned number counts.

\begin{figure}[h]
\vspace{4mm}
\epsfxsize=82mm
{\hfill
\epsfbox{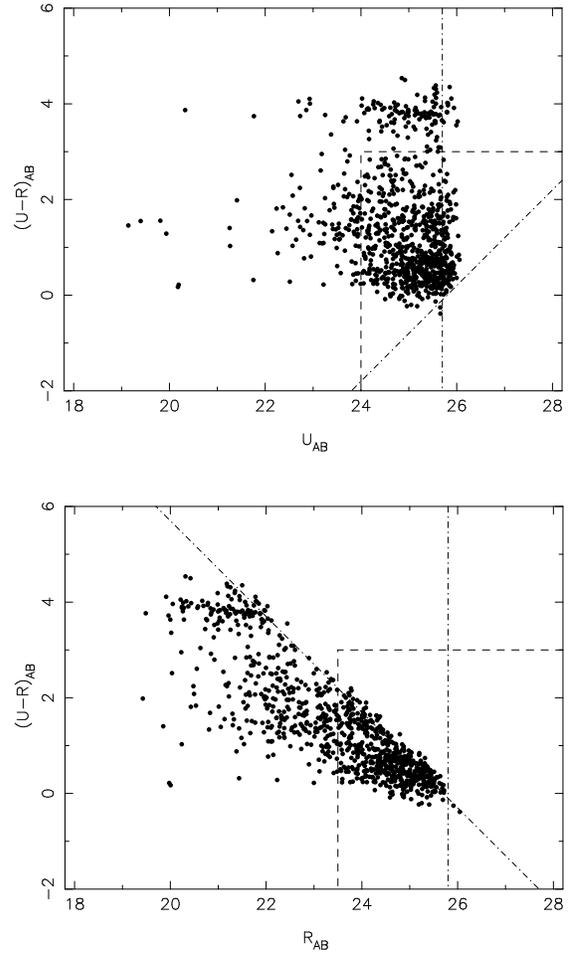}
\hfill}
\epsfverbosetrue
\caption{\small Colour-magnitude diagrams for all 875 matched
objects. Top and bottom shows selection of mask objects in the $U$ and $R$
band respectively by the criteria $(U-R)_{\rm AB}>3$ and $U_{\rm AB}<24$,
$R_{\rm AB}<23.5$ (dashed lines). The $3\sigma$ detection limits
$U_{\rm AB}=25.7$ and $R_{\rm AB}=25.8$ are shown in both plots
by the dot-dashed lines.}
\label{col_mag}
\end{figure}

Figure \ref{col_mag} shows the colour-magnitude plot of $(U-R)_{\rm
AB}$ versus the $U_{\rm AB}$ and $R_{\rm AB}$ magnitudes for the 875
matched objects. By matching our sample with the redshift measurements
of the field of CL0024$+$1654 by Czoske et al. (2001b), we have found
that a large fraction of cluster members can be immediately discarded
by removing objects with $(U-R)_{\rm AB}>3$.  The noticeable lack of
objects seen in the vicinity of $(U-R)_{\rm AB}\simeq 3$ reflects a
minimum in the bi-modal $(U-R)$ distribution of cluster galaxies
identified in the Czoske et al. sample by the criterion
$0.388<z<0.405$.  Applying a further selection $R_{\rm AB}<22$ to
avoid incompleteness in both the Czoske et al. sample and our data, we
find that 65\% of identified cluster objects lie at $(U-R)_{\rm
AB}>3$.  In addition, we find that 12\% of objects brighter than
$R_{\rm AB}=22$ with $(U-R)_{\rm AB}>3$ are foreground galaxies and
only 2\% of objects in this selection are background objects. In
summary, the selection $(U-R)_{\rm AB}>3$ very efficiently removes
cluster galaxies, at least up to $R_{\rm AB}=22$, discarding a very
small fraction of background sources.

\begin{figure}[h]
\vspace{4mm}
\epsfxsize=82mm
{\hfill
\epsfbox{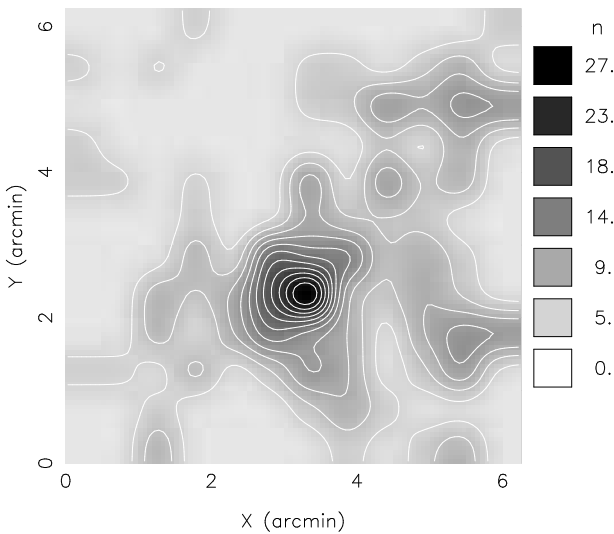}
\hfill}
\epsfverbosetrue
\caption{\small Number density (n=number/square arcmin) of cluster 
galaxies selected either by $(U-R)_{\rm AB}>3$ or redshift.}
\label{cluster_2d_no_dens}
\end{figure}

\begin{figure}
\vspace{4mm}
\epsfxsize=82mm
{\hfill
\epsfbox{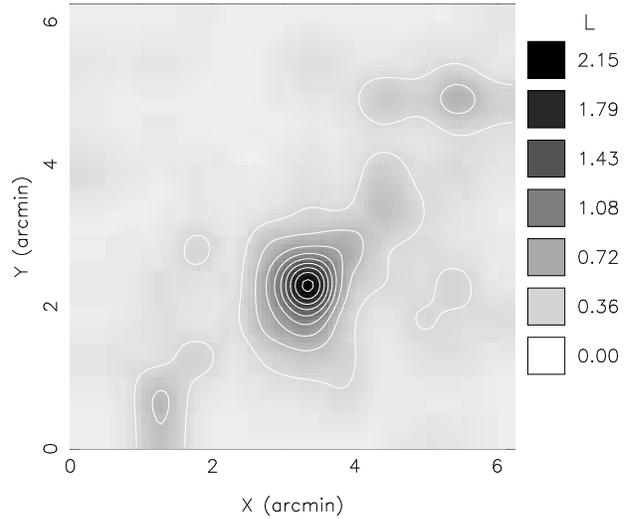}
\hfill}
\epsfverbosetrue
\caption{\small Distribution of light (surface flux density L
converted to the restframe $B$ band and expressed in units of
$10^{13}{\rm L}_\odot {\rm Mpc}^{-2}\,h^2$) from cluster galaxies
selected either by $(U-R)_{\rm AB}>3$ or redshift. Contours are
separated by equal flux intervals.}
\label{clust_gal_light}
\end{figure}

\begin{figure}
\vspace{4mm}
\epsfxsize=82mm
{\hfill
\epsfbox{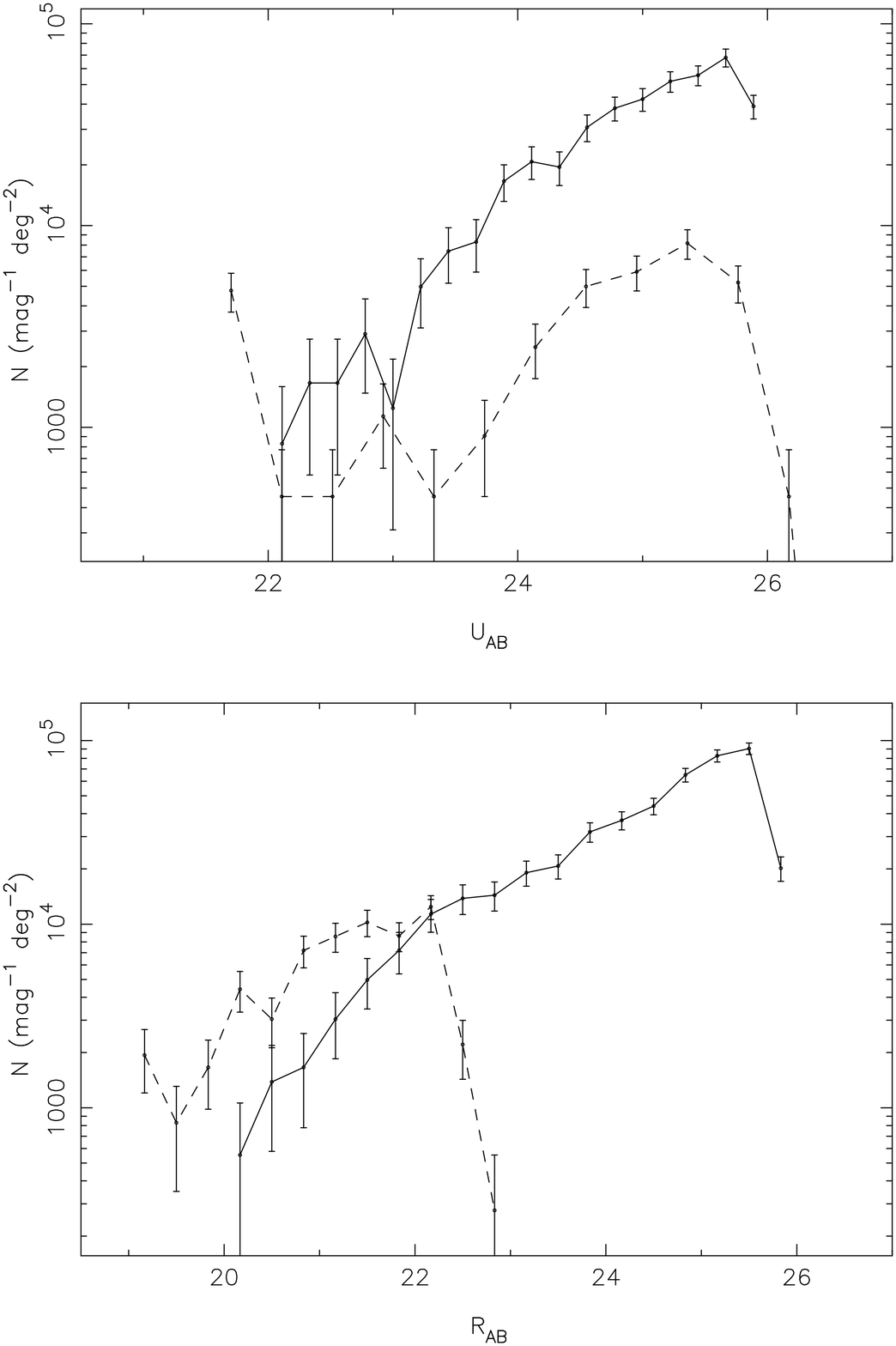}
\hfill}
\epsfverbosetrue
\caption{\small Number counts as a function of magnitude for the $U$
band (top) and $R$ band (bottom). Plotted are counts of galaxies in
the background sample (solid) and cluster galaxies
(dashed). Errors account for shot noise only.}
\label{num_counts}
\end{figure}

In Figure \ref{cluster_2d_no_dens}, we plot the number density of
cluster galaxies defined as objects with colour $(U-R)_{\rm AB}>3$ or
by their redshift for $(U-R)_{\rm AB}<3$. Figure \ref{num_counts}
shows how the number counts of galaxies meeting these criteria vary
with magnitude in the $U$ and $R$. In Figure
\ref{clust_gal_light} we plot the distribution of light from these
selected cluster galaxies. The surface flux density shown in this plot
is scaled to the restframe $B$ band for easy comparison of the mass-to-light
ratio we calculate later with other authors. This scaling was carried
out using the K and colour corrections presented in Fukugita et
al. (1995). Both distributions are compared with the reconstructed
cluster mass map in Section \ref{sec_2Dmass}. The fact that both
distributions are relatively concentrated around the known cluster
centre indicates that we have reliable selection criteria for the
(bright) cluster members.

Our foreground and cluster objects are therefore chosen as a
combination of objects identified in the Czoske et al. sample with
$z<0.405$ as well as those satisfying $(U-R)_{\rm AB}>3$ and an
additional $R_{\rm AB}<23.5$ for the $R$ band sample or an additional
$U_{\rm AB}<24.0$ for the $U$ band sample. The choice of $R$ and $U$
limits here are chosen to be optimal in the sense that they select as
many background objects as possible while preventing too high a degree
of contamination from cluster and foreground objects (see Section
\ref{sec_cluster_lf}). Extracted parameters such as size, ellipticity
and orientation of the cluster and foreground objects are used to
generate an obscuration mask as Section \ref{sec_mask} details.

Rather than perform our analyses on the associated object catalogue
(ie. containing only objects with paired $U$ and $R$ mags) which would
miss faint objects detected in only either the $U$ or $R$ band, we
take our background source list for each band from the individual
complete $U$ and $R$ catalogues.  To obtain our source list, we remove
objects from each complete catalogue identified as cluster members and
foreground objects using the colour-magnitude and redshift criteria
above.

For our $U$ band break analysis in Section \ref{sec_U_band_pop},
removal of contaminants causes a slight dilemma.  Although we would
like our $U$ band sample to be contamination free so that any lensing
signal present is maximised, we also wish to remain consistent with
existing $U$ band studies to prevent our search for the break from
being biased in some way. It turns out that since the Czoske et al.
redshift survey extends to only relatively shallow $U$ band
magnitudes, we only identify and remove 5 foreground objects which
fall within the colour-magnitude selection criteria for our background
sample. We have therefore introduced only a negligible inconsistency
with other $U$ band number count studies and yet have removed all we
can in terms of known foreground objects.  We know that there are more
foreground objects in our background sample than we have been able to
remove (see Section \ref{sec_fgnd_contam}) so we are forced to allow
for these as a contamination error in our lensing analysis. As far as
removal of objects with $(U-R)_{\rm AB}>3$ is concerned, this simply
removes the majority of surplus cluster galaxies which aren't present
in published number count studies anyway.

In the $R$ band, this does not pose a problem. We do not have to
ensure that our $R$ band data is consistent with any other sample
since our depletion is normalised to counts near the edge of the field
rather than independent measurements. Removal of as many foreground
and cluster objects as possible is therefore the ideal. As with the
$U$ band sample, we can not achieve this fully so must resort to
treating the remainder as a source of error in the mass
reconstructions.

After removal of cluster and foreground objects from our full $U$ and
$R$ catalogues, we find a total of 863 background galaxies remaining
in the $U$ and 1367 in the $R$. Figure \ref{num_counts} plots the
number counts of both these samples as a function of magnitude. Notice
how the $U$ band counts are significantly steeper than in $R$. All
subsequent magnification analysis will be performed on these two
catalogues.

\subsection{Contamination of background source samples}
\label{sec_cluster_lf}

We discussed in the previous section that $\sim65\%$ of cluster
galaxies lie at $(U-R)_{\rm AB}>3$ for $R_{\rm AB}<22$. For the $R$
band sample, this does not provide a useful means of removing cluster
galaxies as the background source selection limit of $R_{\rm AB}\geq
23.5$ prevents inclusion of objects with $(U-R)_{\rm AB}>3$ anyway due
to the $U$ band detection limit of $U_{\rm AB}= 25.7$. By including
objects from the full $R$ band catalogue, faint objects with
$(U-R)_{\rm AB}>3$ will inevitably fall into the background $R$
sample. Additionally, the cluster galaxies which lie at $(U-R)_{\rm
AB}<3$ will contaminate both $U$ and $R$ samples along with any
foreground galaxies. This contamination must be quantified.

\subsubsection{Foreground galaxy contamination}
\label{sec_fgnd_contam}

To estimate the foreground galaxy contamination expected within our
chosen magnitude ranges, we use the luminosity functions measured by
the CNOC2 field galaxy redshift survey (Lin et al. 1999).  The CNOC2
survey is ideal for our purposes being the largest intermediate
redshift survey to date with multicolour $UBVRI$ photometry.

The number of objects $N$ which exist within the apparent magnitude range
$m_1<m<m_2$ and the redshift range $z_1<z<z_2$ can be estimated as
\be
\label{eq_ngal_integral}
N=\int^{z=z_2}_{z=z_1}\rmd V(z)\int^{M_2(z)}_{M_1(z)}\phi(M)\,\rmd M 
\ee
where $\rmd V(z)$ is the comoving volume element, $\phi(M)$ is the
luminosity function and the absolute magnitudes $M_1(z)$ \& $M_2(z)$
correspond to the apparent magnitudes $m_1$ and $m_2$ at a redshift
$z$. The fraction of objects which lie closer than a redshift
$z_f$ in a sample of galaxies observed within the 
magnitude range $m_1<m<m_2$ is therefore given by,
\be
\label{eq_forg_contam}
\frac{N(m_1<m<m_2,z<z_f)}{N(m_1<m<m_2)}.
\ee
The denominator here is the total number of galaxies within
$m_1<m<m_2$ integrated over all redshifts.  Using this equation, we
estimate the fraction of galaxies within our $U$ and $R$ background
samples lying at $z<0.405$, allowing for an early+intermediate mixed
galaxy K-correction (Fukugita et al. 1995) for the magnitudes in
equation (\ref{eq_ngal_integral}).


We use the early+intermediate (Cousins) $R$ and (Cousins) $U$
luminosity functions from the CNOC2 survey for calculation of $N$ in
equation (\ref{eq_ngal_integral}). These predict that the
contamination due to foreground galaxies is 3\% in the $U$ band sample
and 2\% in the $R$ band assuming an EdS cosmology.  For the case
$\Omega=0.3$, $\Lambda=0.7$, these fractions drop only very
slightly. We find that this result is insensitive to the
choice of K-correction mix.

\subsubsection{Cluster contamination and the cluster luminosity function}
\label{sec_cluster_contam}

To determine the cluster contamination fraction, we need to
be able to predict the number of cluster members expected within
the magnitude range spanned by the selection magnitude and the detection
limit in both bands. This requires knowledge of the cluster luminosity
function (CLF) for CL0024$+$1654.

We fit our own CLF to the cluster counts determined by matching with
the Czoske et al. cluster members (see Section \ref{sec_obsel}) in
both the $U$ and $R$. We use the likelihood method of Sandage Tammann
\& Yahil (1979) and check the consistency of our results with several
published CLFs. In the $R$ band, we compare with three recent studies
of the composite CLF (CCLF): Paolillo et al. (2001) who construct a
CCLF from 39 Abell clusters, Piranomonte et al. (2000) whose CCLF is
constructed from 86 Abell clusters and Garilli et al. (1999) who use
65 Abell and X-ray selected clusters to calculate a CCLF.  All three
studies use observations in the Gunn $r$ filter so we must apply a
K-colour-correction for the absolute magnitude conversion $M_{R_{\rm
AB}} \rightarrow M_r$. Since $\sim 60-70\%$ of galaxies within
clusters are E/S0 types (Dressler et al. 1997; see also Smail et
al. 1997 for specific case of CL0024$+$1654), we calculate this
correction using an elliptical galaxy spectrum taken from Kinney et al
(1996).

In the $U$ band, publications on CLFs are rare and there are no known
$U$ band CCLFs to date. Applying K-colour-corrections to convert
redder-band magnitudes to the $U$ band when the galaxy type mix is not
known accurately is not reliable. We therefore choose the most heavily
studied cluster in the $U$ band, the Coma cluster, and compare our fit
with the CLF from the recent study by Beijersbergen et al. (2001).  In
a similar manner to the $R$ band, we apply a correction to take
absolute $U_{\rm AB}$ magnitudes at $z=0.39$ to rest-frame absolute
Cousins $U$ magnitudes.

\begin{figure}
\vspace{4mm}
\epsfxsize=82mm
{\hfill
\epsfbox{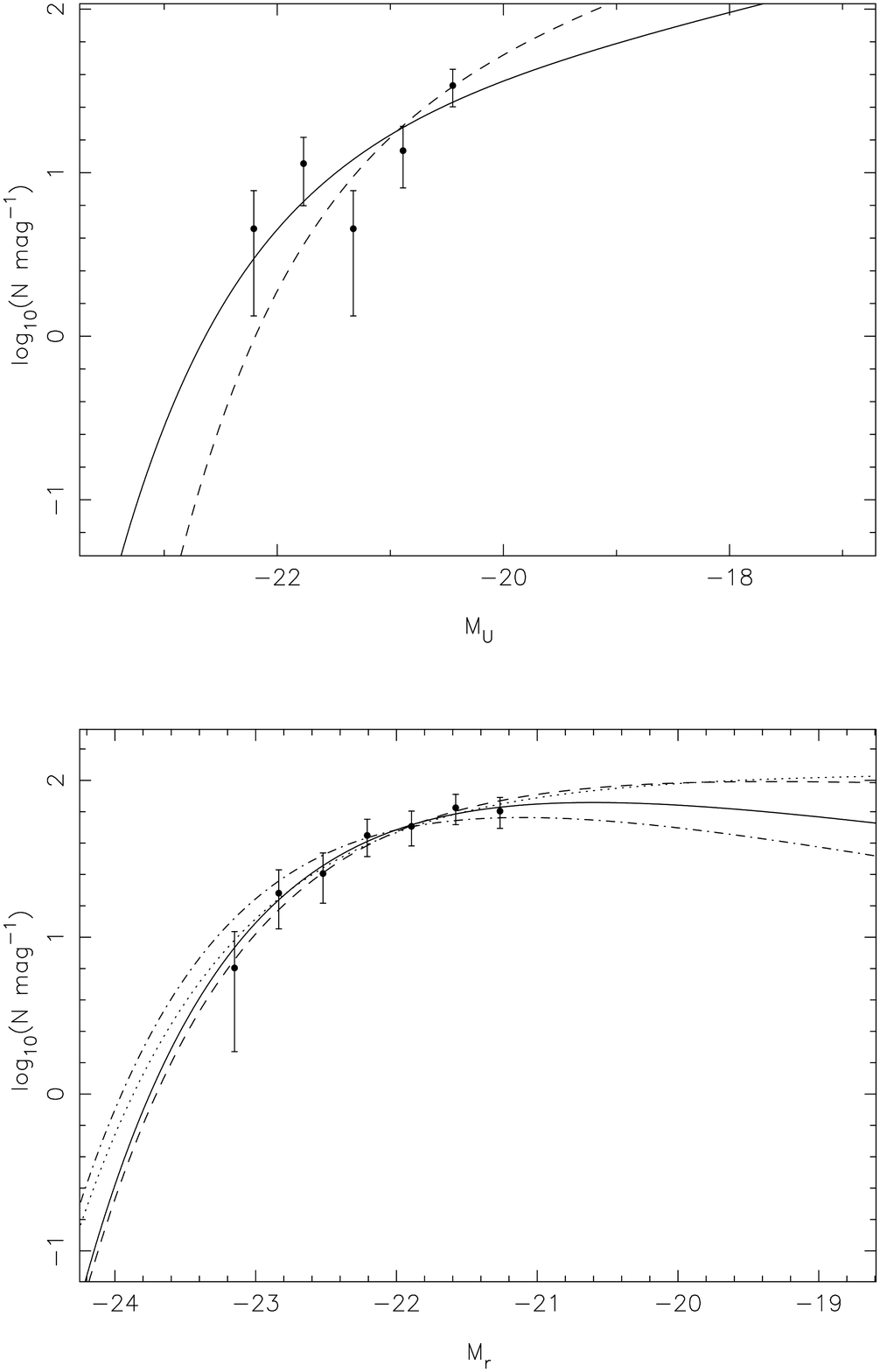}
\hfill}
\epsfverbosetrue
\caption{\small Luminosity functions of CL0024$+$1654 galaxies.  {\em
Top}: $U$ band Schechter LF from Beijersbergen et al. (2001) for the
Coma cluster (dashes) and the best fit LF determined in this work
(solid).  {\em Bottom}: $r$ band CCLFs for rich clusters from Paolillo
et al. (2001, dots), Piranomonte et al. (2000, dashes), Garilli et al.
(1999, dot-dashes) and the best fit from this work (solid).}
\label{cluster_LF}
\end{figure}

Figure \ref{cluster_LF} plots the various CLFs/CCLFs, all of which are
described by a Schechter function (Schechter 1976) with parameters
given in Table \ref{tab_schechter_pars}. For the $r$ band CCLF from
each study, we have taken Schechter parameters corresponding to the
rich subsample of clusters for consistency with CL0024$+$1654.  All
luminosity functions are normalised to our cluster counts which are
superimposed in both plots. Absolute magnitudes assume
$q_0=h_0=0.5$. For correct normalisation, we must ensure that the
Czoske et al. matched sample for each band does not suffer from
incompleteness. We therefore limit the matched $U$ band counts to the
magnitude range $21.8<U_{\rm AB}<24.0$ ($-22.4 < M_U < -20.2$) and the
matched $R$ band counts to $19.9<R_{\rm AB}<22.1$ ($-23.3 < M_r <
-21.1$).

\begin{table}
\vspace{4mm}
\centering
\begin{tabular}{lcccc}
\hline
CLF/CCLF & Band & $M_*$ & $\alpha$ & C (\%) \\
\hline
Paolillo CCLF & $r$ & -22.2 & -1.01  & 15 \\
Piranomonte CCLF & $r$ & -22.0 & -0.91  & 13 \\
Garilli CCLF & $r$ & -22.1 & -0.60  & 5 \\
This work CLF & $r$ & -21.9 & -0.70  & 19 \\
Beijersbergen CLF & $U_c$ & -19.4 & -1.54  & 8 \\
This work CLF & $U_c$ & -21.5 & -1.41  & 9 \\
\hline 
\end{tabular}
\caption{Schechter parameters of the CLF/CCLFs used for estimation of
cluster contamination, C, expressed as the percentage of objects
expected to be cluster members within $23.5 < R_{\rm AB} < 25.5$ for
$r$ or $24.0 < U_{\rm AB} < 25.5$ for $U_c$. Note that the contamination
fractions calculated from the CLFs of this work are
weighted averages (see text).}
\label{tab_schechter_pars}
\end{table}

Integrating over each of the CLFs/CCLFs, we are able to calculate the
number of cluster galaxies expected. For the published luminosity
functions presented here, this is a straight-forward integral.  For
our own fitted CLFs however, we incorporate the uncertainty arising
from their fit to obtain a more fair level of contamination. We
calculate the number of cluster galaxies as
\be
n_c=A\left<\phi^*(M_*,\alpha)\int^{M_b}_{M_a}\rmd M 
\phi(M;M_*,\alpha)\right>_{M_*,\alpha}
\ee
where the angular brackets denote averaging over each realisation of
$M_*$ and $\alpha$ weighted by the corresponding probability from the
CLF fit. $A$ is a normalisation constant calculated as the inverse of
the sum of these probabilities. The absolute magnitudes $M_a$ and
$M_b$ correspond to the apparent magnitude ranges $23.5 < R_{\rm AB} <
25.5$ for $r$ or $24.0 < U_{\rm AB} < 25.5$ for $U_c$. These ranges
are chosen to extend slightly less deep than the $3\sigma$ detection
limits, allowing comparison with a more complete background sample to
give a more accurate estimation of contamination.

Within the appropriate magnitude ranges given above, the last column
of Table \ref{tab_schechter_pars} lists the predicted number of
cluster galaxies as a percentage of the total number of galaxies
detected. In $R$, the estimated contamination from the published
luminosity functions ranges between 5\% and 15\%. The weighted average
predicted contamination from our own $R$ band CLF is higher at 19\%.
In the $U$ band, the contamination levels are not as high since the
known Czoske cluster galaxies have been discounted in the
calculation. We find that the Coma CLF predicts that 8\% of objects in
our background sample are cluster objects, compared with our CLF which
gives a weighted average of 9\%.

The total contamination predicted from both foreground and cluster
galaxies therefore has a relatively wide spread in the $R$ band of 7\%
to 21\% compared to between 11\% and 12\% in $U$. Unfortunately, none
of the published CLFs/CCLFs presented here extend sufficiently deep to
be able to reliably predict the number of cluster objects in
CL0024$+$1654 at the faint end of our sample. In addition, some
CLF studies (eg. Driver et al. 1994; Wilson et al. 1997) indicate
the possibility of steeper faint end slopes which would increase the
fraction of cluster contaminants. In light of this, we therefore
assume the generous contamination levels predicted from our fitted
CLFs in the analysis which follows in this paper.

Clearly, the contamination is dominated by the cluster members. This
means that the variation in contamination across the field of view is
governed by their density profile. Since this increases toward the
centre of the cluster and since the background number counts fall off
toward the centre where the magnification is strongest (see Section
\ref{sec_R_band_fits}) the observed number density of objects in the
background sample is affected most by contamination at small
radii. This distribution will have a strong fall-off as one moves away
from the cluster centre.  Fortunately, this reduces the impact of the
contamination on the cumulative mass measurements (Section
\ref{sec_cum_mass}) at large radii where the strongest mass
contribution comes from.

An approximate contamination profile can be obtained
by assuming that the cluster mass distribution and hence the number
density of galaxies is described sufficiently accurately (for this
purpose at least) by an isothermal sphere (see Section
\ref{sec_R_band_fits}). The radial dependence of the error on
the surface mass density, $\kappa$, can thus be written
\be
\label{eq_contam_error}
\sigma_\kappa(r) = \frac{k}{2nr(\beta-1)\mu^{\beta}(r)}
\ee
where $n$ is the background number density in the absence of lensing,
$\mu(r)$ is the magnification given by equation
(\ref{eq_mag_iso_sphere}) and $\beta$ is the number count slope (see
Section \ref{sec_sngl_powlaw}).  We have approximated the number
density profile of the cluster galaxies as $k/r$ where $k$ is set by
normalising to the contamination fractions given earlier
in this section.

In all reconstructions of $\kappa$ found later in this paper, we add
the error given by equation (\ref{eq_contam_error}) in quadrature to
the sum of the other errors, including the error due to foreground
galaxy contamination. The effect of contamination on the search for
the $U$ band break is discussed separately in Section
\ref{sec_U_band_pop}.

\subsection{Obscuration masks}
\label{sec_mask}

To account for obscuration of the background sky by cluster members
and foreground objects, we create a mask. Without giving consideration
to obscuration, the ratio of observed objects to expected objects used
later in our analysis would be biased to lower values. This is
especially true of bins near the cluster centre where relatively heavy
obscuration occurs due to large central cluster galaxies.

\begin{figure}
\vspace{4mm}
\epsfxsize=82mm
{\hfill
\epsfbox{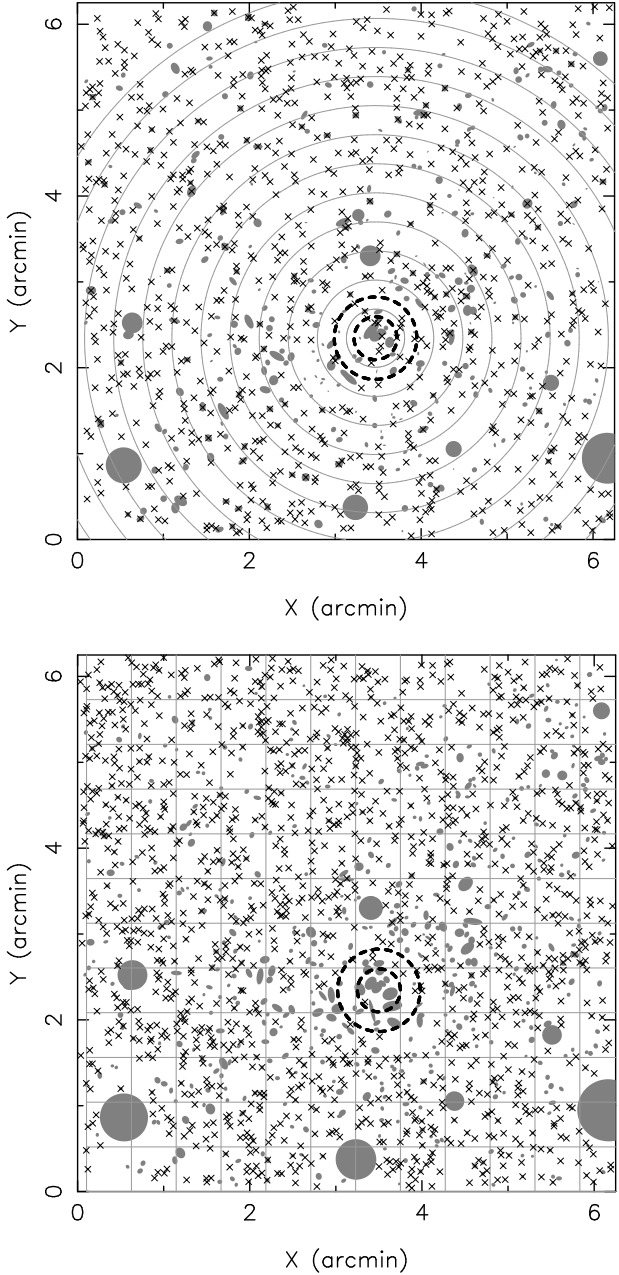}
\hfill}
\epsfverbosetrue
\caption{\small {\em Top}: $U$ band mask (grey ellipses) with source
positions (black crosses) and annular bins used in Section
\ref{sec_R_band_fits} and \ref{sec_U_band_pop}. {\em Bottom}:
Corresponding $R$ band mask with grid used for 2D maps in Section
\ref{sec_2Dmass}. Observed arcs lie along the outer
heavy dashed circle in both plots. The inner dashed circle shows
the critical line determined from the isothermal
sphere model fit to the $R$ band number counts
(see Section \ref{sec_R_band_fits}).}
\label{masks}
\end{figure}

Figure \ref{masks} shows the obscuration mask produced using the
SExtracted parameters of mask objects selected in Section
\ref{sec_obsel}.  The top half shows the $U$ band mask (grey
ellipses), $U$ band background source positions (black crosses) and
the position of the annular bins used in Section \ref{sec_R_band_fits}
and Section \ref{sec_U_band_pop} to bin counts radially.  The lower
half of Figure \ref{masks} shows the corresponding $R$ band mask and
background sources with the grid used in Section \ref{sec_2Dmass} for
2D binning. Note the horizontal phase of the bins with respect to the
field of view to allow placement of bin $(7,5)$ directly over the
cluster centre, aligning it with the area encompassed by the critical
line.

\section{Mass reconstruction theory}
\label{sec_recon_theory}

In this section, we outline the process used for the determination of
magnification from number counts. We then detail the three different
methods used to transform from magnification to surface mass density.

\subsection{Lens magnification: Single power-law number counts}
\label{sec_sngl_powlaw}

A background population of galaxies whose integrated
number counts follow the standard power-law $n(<m)\propto
10^{0.4\beta m}$ will be observed under a lens magnification factor $\mu$ 
to have the number count $n'(<m)$ given by (BTP, T98);
\be
\label{eq_mag_eqn}
n'=n\mu^{\beta-1}(1+\delta_{nl})=\lambda(1+\delta_{nl}).
\ee 
The quantity $\delta_{nl}$ accounts for perturbations in $n$ due to
non-linear clustering and we define $\lambda$ as the expected number
of sources in the absence of clustering. We model the fluctuation
$\delta_{nl}$ with a lognormal distribution and combine this with the
additional uncertainty due to shot noise to give a Poisson-lognormal
distribution (eg. Coles \& Jones 1991; BTP). Unlike a Gaussian
distribution, the lognormal distribution accounts for non-linear
clustering of the background probed by our small field of view and is
positive-definite.

The Poisson-lognormal distribution can be defined as a compound distribution
formed from a Gaussian distribution ${\rm G}$ of a linearised field 
and a Poission distribution ${\rm D}$ as 
\be
\label{eq_like_mag}
{\rm P_{LN}}[n|\lambda(\mu)]=\int^{\infty}_{-\infty}\rmd \delta \,
{\rm D}[n|\lambda e^{\delta-\sigma^2/2}]{\rm G}(\delta).
\ee
Here, $\delta$ is the linear density fluctuation which relates to
the non-linear fluctuation as $1+\delta_{nl}=e^{\delta-\sigma^2/2}$
with $\sigma$ the linear clustering variance. Following the method of
T98 but applied to the $R$ band correlation function of Hudon \&
Lilly (1996), we calculate the non-linear clustering
variance. Averaging the angular correlation function over a 
circular area of radius $\theta$ gives,
\be
\sigma_{nl}^2=1.5\times 10^{-2}z^{-1.8}(\theta/1')^{-0.8}.
\ee
The linear clustering variance is calculated from the non-linear
variance using $\sigma^2=\ln(1+\sigma_{nl}^2)$.

Equation (\ref{eq_like_mag}) gives the probability distribution for
the lens magnification in a given bin containing $n$ galaxies. This
directly gives the most probable magnification for that bin with its
associated error taken as the width of the distribution. The integral
in equation (\ref{eq_like_mag}) must be evaluated numerically.

\subsection{Lens magnification: Dual power-law number counts}
\label{sec_dual_slope_mag}

In Section \ref{sec_Uband_slopes}, we discuss evidence which suggests
the presence of a break in the $U$ band number count slope at faint
magnitudes. Section \ref{sec_R_band_char} also highlights the
possibility of a break in the $R$ band counts.  To account for the
effect this has on the magnification calculated from number count
depletion, we modify equation (\ref{eq_mag_eqn}) to incorporate a
second slope applicable beyond some break magnitude $m_b$.

Writing the observed {\em differential} number counts as a dual power-law,
\be
n(m)=\left\{\begin{array}{lll}
a10^{0.4\beta_1 m} & , & m<m_b \\
b10^{0.4\beta_2 m} & , & m \geq m_b \\ \end{array} \right.
\ee
where the normalisation coefficients $a$ and $b$ are constrained by
continuity, the integrated number counts up to some limiting
magnitude $m_l > m_b$ for $\mu\geq 1$ can be expressed as
\ba
\label{eq_mag_eqn_dual}
n'(<m_l)&=& \mu^{-1}\left[n_1(<m_b)+\right. \nn
& &\left. n_2(<m_l)\mu^{\beta_2}-n_2(<m_b)\right].
\ea
Here, $n_1$ and $n_2$ denote number counts integrated up to a
limiting magnitude over a constant slope of $\beta_1$ and $\beta_2$
respectively. The clustering term which features in equation
(\ref{eq_mag_eqn}) has been omitted here for
clarity but is used in our analysis later.

If the limiting magnitude of observations matches the break
magnitude then
\be
n_2(<m_l)=\frac{\beta_1}{\beta_2}n_1(<m_l).
\ee
Equation (\ref{eq_mag_eqn_dual}) can thus be simplified to
\be
\label{eq_mag_eqn_dual_simple}
n'=n\mu^{-1}\left[1+
\frac{\beta_1}{\beta_2}\left(\mu^{\beta_2}-1\right)\right], 
\quad (\mu \geq 1)
\ee
where $n=n_1(<m_l)$ here, is the unlensed surface number count density
observed, for example, at the edge of the field of view. If
$\beta_1=\beta_2$ and hence there is no break, equation
(\ref{eq_mag_eqn_dual_simple}) becomes equation
(\ref{eq_mag_eqn}). For the case when $\mu < 1$, equation
(\ref{eq_mag_eqn}) must be reverted to.

\subsection{Local mass estimator}
\label{sec_est}

The simplest means of arriving at a mass estimate given a measurement
of magnification is by assuming a local relationship between the
convergence and shear. Such a relationship can be derived once
a model is chosen to describe the mass distribution.
Since magnification depends on the convergence, $\kappa$, and shear,
$\gamma$, as
\be
\label{eq_mag}
\mu=\left|(1-\kappa)^2-\gamma^2\right|^{-1},
\ee
this gives a directly invertible relation for $\mu$ in terms of
$\kappa$ and hence allows $\kappa$ to be estimated.

For simplicity, we use the `parabolic estimator' suggested by the
cluster simulations of van Kampen (1998) which relates $\gamma$
to $\kappa$  via
\be
\gamma=\left|1-c\right|\sqrt{\kappa/c} \,\, .
\ee
As in T98, we adopt the value $c=0.7$ corresponding to a profile which lies
between that of a homogeneous sheet of matter ($\kappa=constant$) and
an isothermal profile ($\kappa\propto r^{-1}$, $r$ the distance from
the cluster centre).  The magnification can therefore be written
\be
\label{eq_mag_par}
{\cal P}\mu^{-1}=\left[(\kappa-c)(\kappa-1/c)\right],
\ee
where ${\cal P}=\pm 1$ accounts for image parity on either side of the
critical line implied by the modulus in equation (\ref{eq_mag}).
Rearranging this equation for $\kappa$ gives
\be
\kappa=\frac{1}{2c}\left((c^2+1)-{\cal S}
\sqrt{(c^2+1)^2-4c^2(1-{\cal P}\mu^{-1})}\right).
\ee
The second parity ${\cal S}$ is due to the parabolic nature of
equation (\ref{eq_mag_par}) which permits a second critical line and
thus higher values of $\kappa$. Since there is no evidence for the
existence of a double critical line in CL0024$+$1654, we will adopt
${\cal S}=+1$ throughout this paper.

Since the parabolic estimator requires only local $\kappa$ and
$\gamma$, it finds application to both radial and 2D magnification
distributions in this paper.

\subsection{Self-consistent axi-symmetric mass estimator}
\label{sec_axi_symm_est}

The second method which we use to estimate cluster mass is the
so-called self-consistent axi-symmetric mass estimator introduced by
T98. The non-local nature of this estimator allows the magnification
equation (\ref{eq_mag}) to be solved for a self-consistent $\kappa$
and $\gamma$ radial profile. Although the estimator is valid only for
axi-symmetric mass distributions, it can be applied to data binned in
any self-similar set of contours centred on the peak of the mass
distribution.

In a given annular bin $i$, the shear can be expressed as
\be
\label{eq_axi_gam}
\gamma_i=\left|\kappa_i-\overline{\kappa}_i\right|
\ee
where $\kappa_i$ is the convergence in the bin and $\overline{\kappa}_i$
is the convergence averaged over the area interior to and including
the bin.  Substituting this into equation (\ref{eq_mag}) gives for the
magnification in bin $i$
\be
\label{eq_mag_axi}
{\cal P}\mu^{-1}_i=(1-\overline{\kappa}_i)(1-2\kappa_i+\overline{\kappa}_i).
\ee
Dividing $\overline{\kappa}_i$ into two terms, one for bin $i$
and the other, which we denote $\eta_{i-1}$, for all interior bins, so
that
\be
\label{eq_avkap_axi}
\overline{\kappa}_i=\eta_{i-1}+\frac{2}{i+1}\kappa_i
\ee
allows rearrangement of equation (\ref{eq_mag_axi}) to give
\ba
\label{eq_kap_axi}
\kappa_i &=& \frac{(i+1)}{4i}\{i+1-(i-1)\eta_{i-1}- \nn 
& & {\cal S} [(i-1-(i+1)\eta_{i-1})^2+4i{\cal P} \mu_i^{-1}]^{1/2}\}.
\ea
The parities ${\cal P}$ and ${\cal S}$ have the same function
as in Section \ref{sec_est}. 

Using equation (\ref{eq_kap_axi}), $\kappa$ is calculated
iteratively. The only freedom is choice of $\eta_0$ which as equations
(\ref{eq_axi_gam}) and (\ref{eq_avkap_axi}) show, is $\gamma_1$, the
shear in the first bin. To avoid non-physical solutions,
$\gamma_1^2\geq{\cal P}\mu^{-1}_1$ must be enforced leaving only a
sensible range of values. As T98 discuss, the overall mass and shear
profile obtained with the self-consistent axi-symmetric estimator
proves to have only a minor sensitivity to the choice of $\gamma_1$.

\subsection{Self-consistent 2D mass estimator}

The final reconstruction method we apply in this paper is that
discussed by Dye \& Taylor (1998). By pixellising the image into 
a rectangular grid of pixels, the components of the shear in any pixel
$n$ can be expressed as
\be
\label{eq_gam_pix}
\gamma_i^n=D_i^{mn}\kappa_m, \quad i=1,2
\ee
where summation over index $m$ is implied for all $N$ pixels
on the grid. The matrices $D_1$ and $D_2$ are simple geometrical
functions of the different combinations of positions of pixels
$m$ and $n$ (see Dye \& Taylor 1998 for details).

Writing equation (\ref{eq_mag}) in its pixellised form and
substituting the expression for shear in equation 
(\ref{eq_gam_pix}) gives the vector equation
\be
\label{matrix_eqn}
\bone-2\kb+\kb \bG \kb^{\rm t} - {\cal P}\mub^{-1} = 0
\ee
where ${\cal P}$ is again the image parity from Section \ref{sec_est},
$\mub^{-1}$ is the $N$-dimensional vector of pixellised inverse
magnification values, $\kb^{\rm t}$ is the transpose of the
vector $\kb$ of pixellised convergence values and $\bone$ is
the vector $(1,1,1,...)$. $\bG$ is an $N\times N\times N$ matrix
with the elements
\be
G_{pqn} = \delta_{pn}\delta_{qn} - D_1^{pn} D_1^{qn} - 
D_2^{pn} D_2^{qn}
\ee
where $\delta_{ij}$ is the Kr\"{o}necker delta, and summation occurs
only over indices $p$ and $q$ in equation (\ref{matrix_eqn}).

We solve equation (\ref{matrix_eqn}) for $\kappa$ using a hybrid Powell
method provided by the NAG routine C05PCF.

\section{Cluster Model constraints from Number Counts}
\label{sec_no_count_fitting}

Before application of the theory presented in Section
\ref{sec_recon_theory}, the properties of the background population
must be understood.  Our initial attention is turned toward the $R$
band population of background galaxies since this proves to be the
most suitable for the application of lens magnification owing to its
shallower number count slope. In this section, we use the radially
binned number counts of the sample to fit an isothermal, power-law and
NFW profile.  We defer discussion of the $U$ band sample properties
until Section \ref{sec_U_band_pop}.

\subsection{$R$ band sample characteristics}
\label{sec_R_band_char}

The characteristics of the $R$ band background source galaxy
population must be constrained before we can apply our lensing
analysis to the source counts. Specifically, the unlensed surface
number density $n_R$ and the number count slope must be determined.

T98 estimate the unlensed surface number density from deep number
counts of field galaxies observed independently of their work.  We opt
for the alternative choice here of taking $n_R$ from our field where
the lens effect of the cluster is expected to be small. This
eliminates potential biasing of the reconstructions which arise from
differing completeness characteristics between the lensed field and
the reference unlensed field (Gray et al. 2000). This is particularly
important for the $U$ band break search in Section
\ref{sec_U_break}. In an annular region centred on the cluster and
bounded by the radial limits $120''<r<180''$, we measure $n_R=43\pm2$
arcmin$^{-2}$ in the $R$.  This value is assumed in our subsequent
analysis.

Ideally, this normalising annulus should be further away from the
centre of the cluster than we have chosen. However, as we discuss in
Section \ref{sec_results}, the edge of our field is somewhat noisy due
to the presence of some bright stars.  This causes a drop in the
number counts at larger radii so that normalising here would force a
large over density of counts at medium radii and hence negative
mass. An alternative explanation is that such an overdensity would be
caused by contaminating cluster galaxies.  In fact, it is most likely
that both effects jointly contribute.  Since our reconstructions
include the uncertainty due to cluster contaminants, any bias as a
result of normalising out this effect falls within our error
budget. We therefore expect only a small bias to remain from the
underestimation of background counts at large radii.

As far as the number count slopes are concerned, the $R$ band slope
determined by Hogg et al. (1997) over $21<R_{\rm AB}<25$ is
$\beta_R=0.83$, in agreement with Smail et al. (1995) who measure
$\beta_R=0.80$ over the same magnitude range.  This is consistent with
the slopes in the $V$ and $I$ from the Hubble Deep Field data of
Pert et al. (1998) with $\beta_V\simeq0.9$ and $\beta_I\simeq0.8$
over $23<(V,I)_{\rm AB}<26$. However, over the fainter magnitude range
$26<(V,I)_{\rm AB}<29$, they measure a shallower slope of
$\beta_{V,I}<0.5$ and find the trend that flattening of the number
count slope is more pronounced in shorter wavelength bands. The fact
that this is seen at all in the $I$ band therefore suggests that
flattening most probably also occurs in $R$ at these faint magnitudes.
This is fortified by the more recent $R$ band counts of Metcalfe et
al. (2001) from the William Herschel Deep Field, which suggest a slope
of $\beta_R\simeq 0.6$ fainter than $R_{\rm AB}\simeq 26$.

Our $R$ band observations tantalisingly extend to approximately the
depth where the apparent break occurs. We compare in Section
\ref{sec_R_band_fits} and \ref{sec_cum_mass} the difference between
the results obtained using the single and dual slope models.  In the
analysis hereafter, for the single slope model, we take $\beta_R=0.80$
and for the dual slope model, $\beta_{R1}=0.80$ and $\beta_{R2}=0.5$
with a break magnitude of $R=26$.

\subsection{Magnification profiles}
\label{sec_magprof_R}

In the case of lens magnification, the least biased method
of determining the best fit mass model is to fit depletion
curves to the number count profile. In this way, the lensing signal
is used in its purest form before potential biases are introduced
by calculation of the $\kappa$ profile.

To fit the depletion profile from a given mass model, its lens
magnification must be determined. We choose to fit an isothermal
sphere, a power-law mass model and a NFW profile, all of which have
analytical forms for their magnification. In the case of the
isothermal sphere, the mass profile is completely determined by its
critical radius $r_{crit}$,
\be
\kappa(r)=\frac{r}{2r_{crit}},
\ee
giving a magnification of
\be
\label{eq_mag_iso_sphere}
\mu(r)=\left|1-r_{crit}/r \right|^{-1}.
\ee

The power-law model we choose is that of Schneider, Ehlers \& Falco (1993)
which gives a smooth, non-singular surface mass density distribution
with a $\kappa$ profile given by
\be
\kappa(x)=\kappa_0\frac{1+\xi\,x^2}
{\left(1+x^2\right)^{2-\xi}}.
\ee
Here $\kappa_0$ is the peak surface mass density at the centre, $\xi$
is the power-law slope and $x=r/r_{core}$ with $r_{core}$ the core
radius. For this distribution to remain positive definite at large
radii and so that it is a declining function of radius, the valid
range of slopes is $0<\xi<1$.  At $r\gg r_{core}$, this physically
corresponds to a range of mass models spanning a homogeneous sheet
($\xi=1$) through an isothermal sphere ($\xi=0.5$) all the way up to
$\kappa \propto 1/x^4$ ($\xi=0$). The magnification resulting from
such a profile is (Schneider, Ehlers \& Falco 1993)
\ba
\label{eq_mag_pl}
\mu(x)=& &\left[1-\frac{\kappa_0}{(1+x^2)^{1-\xi}}\right]^{-1} \times \nn
& &\left[1-\frac{\kappa_0}{(1+x^2)^{2-\xi}}
\left\{1+(2\xi-1)\xi^2\right\}\right]^{-1}.
\ea

Finally, the NFW model has a $\kappa$ profile given by (Bartelmann 1996)
\be
\kappa(y)=2\kappa_s\frac{f(y)}{y^2-1}
\ee
where
\be
f(y)=\left\{\begin{array}{lll}
1 - \frac{2}{\sqrt{y^2-1}}\tan^{-1}\sqrt{\frac{y-1}{y+1}} & , & (y>1) \\
1 - \frac{2}{\sqrt{1-y^2}}\tanh^{-1}\sqrt{\frac{1-y}{1+y}} & , & (y<1) \\
0 & , & (y=1) \\
\end{array}\right.
\ee
and $y=r/r_s$. The scale radius, $r_s$, and surface density
normalisation, $\kappa_s$, are free parameters of the model. The
magnification of the NFW model comes from Jacobian determinant of
the lens mapping for axisymmetric lenses (Schneider, 
Ehlers \& Falco 1992);
\be
\mu^{-1}(y)=\left(1-\frac{m(y)}{y^2}\right)
\left[1-\frac{\rmd}{\rmd y}\left(\frac{m(y)}{y}\right)\right]
\ee
where
\be
m(y)=2\int_0^{y}\rmd y' y'\kappa(y').
\ee

\subsection{Number count profile fits}
\label{sec_R_band_fits}

\begin{figure}
\vspace{4mm}
\epsfxsize=82mm
{\hfill
\epsfbox{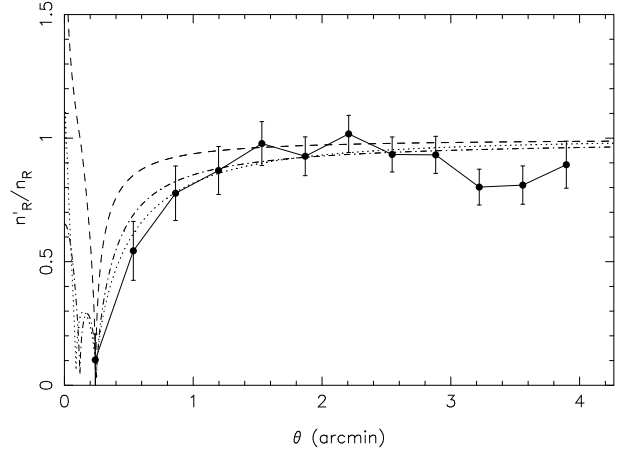}
\hfill}
\epsfverbosetrue
\caption{\small Number of counts as a ratio of expected counts versus
radial distance from cluster centre in the $R$ band.  Error bars
account for the error in $n_R$ and shot noise.  Superimposed are the
isothermal (dashes, $r_{crit}=15''$), power-law (dot-dashes,
$\kappa_0=1.52$, $\xi=0.61$, $r_{core}=11''$) and NFW (dots,
$r_s=1.75'$, $\kappa_s=0.23$) models fit to the first 9 points. }
\label{n_n0_R}
\end{figure}

Using equation (\ref{eq_mag_eqn}) for the single slope model (and
neglecting the clustering term for now) with the three forms for $\mu$
from Section \ref{sec_magprof_R}, we perform a $\chi^2$ fit to the
observed depletion profiles.  Figure \ref{n_n0_R} shows the $R$ band
radial number counts expressed as a fraction of the intrinsic
background counts.  The degree of obscuration by foreground objects is
taken into consideration by adjusting $n_R$ in each bin. The $1\sigma$
error bars plotted account for the uncertainty in $n_R$ and
shot-noise.

Superimposed on the observed $R$ band number counts in Figure
\ref{n_n0_R} are the isothermal, power-law and NFW profiles obtained
from fitting to the first 9 data points.  We discuss in Section
\ref{sec_2Dmass} that the outer profile is affected by noisy features
at the edge of the field of view. The final model fits therefore omit
the last 3 points although we also fit to all 12 points to investigate
the variation in fitted parameters.

For the isothermal profile, a best fit critical radius of
$r_{crit}=15''\pm1''$ is obtained using either the first 9 points or
all 12 points. Note that this is somewhat smaller than the estimate of
$r_{crit}=25''$ in the $R$ band from R01. This discrepancy arises partly
from the different object selection and extraction criteria used by
R01, detailed in Section \ref{sec_the_data}, and also from our
different radial bin width. The error of $\pm1''$ we quote here is
solely the error from the $\chi^2$ fit which does not include any
uncertainty to allow for the choice of binning. Variation in bin width
and also the radius attributed to a given bin adds further error.  In
Figure \ref{n_n0_R}, the radial position of a given bin is taken as
the radius which divides that bin into two equal areas. Since the
isothermal fit is dominated by the radius of the first data point,
adopting different binning strategies affects the fitted value of
$r_{crit}$ substantially. Measuring the variation in fitted values of
$r_{crit}$ with different binnings, we find that a further error of
$\pm 10''$ should be included to give an overall fit of
$r_{crit}=15''\pm 10''$.
 
The observed radius of the large arcs, $r_{arc}=30''$, is also
somewhat larger than our fitted critical radius. One explanation for
this would be that the background population we select lies at a lower
redshift on average than the lensed background galaxy forming the arc
at $z=1.675$ (B00). In fact, knowing the amount of mass contained
within the arcs (see Section \ref{sec_cum_mass}), one finds that a
critical radius of $r_{crit}=15''\pm 10''$ corresponds to a background
population with a mean redshift lying within the range
$0.40<z_{mean}<0.82$ (EdS or $\Omega=0.3$, $\Lambda=0.7$). Note that
although this range appears to permit only quite low mean redshifts,
it depends sensitively on the confidence interval.  For example,
expanding the interval from 68\% quoted above to merely 80\% results
in a range $0.40<z_{mean}<1.19$. Clearly, the large error on
$r_{crit}$ prevents an accurate measurement of $z_{mean}$.

However, considering the redshift survey of Cohen et al. (2000) in the
Hubble Deep Field North region, this indicates a median redshift of
$z_{median}=0.79^{+0.30}_{-0.31}$ in their deepest bin at $R_{\rm
AB}\simeq 24$. Extrapolating their $z_{median}$ versus magnitude plot
by a further $1.5$ mags to coincide with our $3\sigma$ detection limit
and weighting by the expected number counts over this interval, one
obtains an approximate mean redshift expected for our background $R$
sample of $z_{mean}\sim 1.2$. Therefore, although our lens-inferred
measurement of $z_{mean}$ is a little low, it is not significantly
inconsistent.

The choice of bin width of $20''$ means that our ability to constrain
the small scale core radius in the power-law profile is very
limited. In fitting to the power-law, we therefore hold
$r_{core}=11''$ determined by the shear study of CL0024$+$1654 by
TKD which has superior resolution
in the centre of the cluster.  Fitting the remaining two parameters to
the first 9 points then yields $\kappa_0=1.52\pm0.20$ and
$\xi=0.61\pm0.11$ with $1\sigma$ errors accounting for the fit and
binning variation. This compares to $\kappa_0=1.42\pm0.21$ and
$\xi=0.67\pm0.11$ obtained when the remaining 3 points are
included.

Finally, fitting the NFW model to the first 9 points yields the
parameters $r_s=1.75'\pm1.02'$ and $\kappa_s=0.23\pm0.08$.  Including
all 12 points, these become $r_s=2.50'\pm1.32'$ and
$\kappa_s=0.19\pm0.08$. The large errors here reflect a strong
degeneracy between $\kappa_s$ and $r_s$.

Although close to isothermal with a slope of $\xi=0.61$, the power-law
model provides a better fit to the number counts than the isothermal
sphere. The NFW and power-law models are clearly very similar however
the reduced $\chi^2$ of $0.41 \pm 0.53$ from the NFW model (assuming
Gaussian statistics to obtain the error) quantifies the fact that it
is a better fit to the data than the PL model with a reduced $\chi^2$
of $0.62 \pm 0.53$ including the first nine bins. For comparison, the
isothermal sphere model fits to the data with a reduced $\chi^2$ of
$1.33 \pm 0.50$. This agrees with the findings of TKD who concluded
that although the NFW profile predicts too much mass within the inner
arc region, there is little to distinguish this from a power-law
profile at larger radii. In fact, with all 12 bins in the fit, the PL
model fares best with a reduced $\chi^2$ of $1.08\pm0.45$ compared
with that of the NFW model of $1.24\pm0.45$.

Our magnification profile fits prove to be consistent with fits to
shear-derived mass profiles from other work. TKD find that the
mass profile of CL0024$+$1654 is best described by a power-law model
with $\xi=0.57\pm0.02$ and a central surface mass density of
$\Sigma_0=7900\pm 100 h {\rm M}_{\odot} {\rm pc}^{-2}$.  We can
compare this central surface mass density with our estimate of
$\kappa_0$ by integrating our fitted power-law model for $\kappa$ over
the disk enclosed by the observed arcs. Comparing this with the real
projected mass contained within this area from Section
\ref{sec_cum_mass} allows the normalisation of the model to be
calculated. This requires a central surface mass density of
$\Sigma_0=9000\pm 1800 h {\rm M}_{\odot} {\rm pc}^{-2}$, slightly
higher than the TKD estimate but consistent given the error
budget.

Similarly, B00 find that their azimuthally averaged mass profile for
CL0024$+$1654 is close to the NFW prediction for massive clusters with
an overdensity of $\delta_c\simeq 8000$ and $r_s\simeq 400\kpcoh$
(NFW, Ghigna et al. 1998). Using the fact that
$\delta_c=\kappa_s\Sigma_{\rm CR}/(r_s\,\rho_{crit})$ (Bartelmann
1996) where $\Sigma_{\rm CR}$ is the critical lens surface mass
density (see for example Blandford \& Narayan 1992) and $\rho_{crit}$
is the critical density of the Universe, this converts to $\kappa_s
\simeq 0.2$ with a scale radius $r_s\simeq 2'$. The findings of this
paper are therefore in good agreement with the NFW expectation.

Finally, we have also applied the dual slope model to fit the three
profile forms with magnification calculated using equation
(\ref{eq_mag_eqn_dual_simple}). We find that this gives fitted model
parameters within a few percent of those presented above. As expected,
this negligible difference indicates that the lensing signal in the
$R$ band is dominated by galaxies in the $R<26$ magnitude range where
the steeper slope applies, despite flux magnification of objects by
CL0024$+$1654.

\section{Cluster Mass Reconstructions}
\label{sec_results}

In this section, we present the results of the 2D and radial 
mass reconstructions obtained using the $R$ band background
galaxy sample.

\subsection{2D mass maps}
\label{sec_2Dmass}

\begin{figure}
\vspace{4mm}
\epsfxsize=82mm
{\hfill
\epsfbox{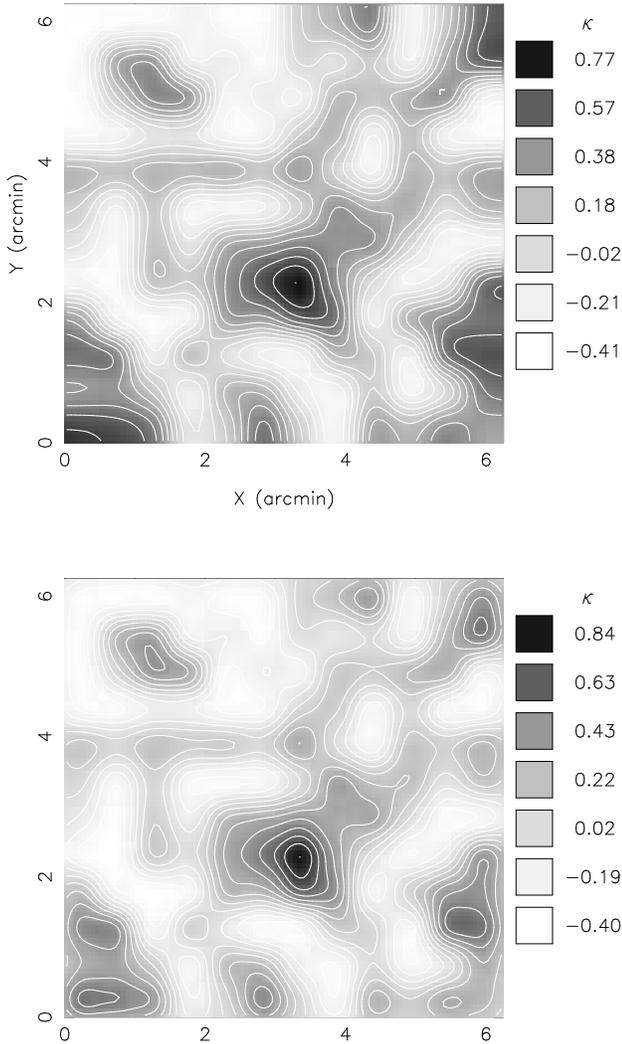}
\hfill}
\epsfverbosetrue
\caption{\small Mass reconstruction from the $R$ band data. {\em Top}: 
Estimated $\kappa$ from parabolic estimator. {\em Bottom}: Iterated 
self-consistent $\kappa$. Contours in both plots are separated by
intervals of $\delta\kappa=0.1$. North is up, East is left.}
\label{2d_kappa_R}
\end{figure}

Using the $12\times 12$ grid of bins shown in Figure \ref{masks} and
the associated $R$ band obscuration mask, source numbers were binned
across the field of view and used to calculate magnifications using
equation (\ref{eq_like_mag}).  Applying both the local estimator and
the self-consistent reconstruction method assuming single slope
counts, we derive the mass maps shown in Figure \ref{2d_kappa_R}.  The
top half of this figure shows the locally estimated $\kappa$ for
comparison with the lower half showing the self-consistent mass
distribution.

Both maps exhibit a very similar mass structure demonstrating that the
local estimator works well. The peak of the distribution in the
self-consistent map is slightly higher than in the estimated map and
is more distinct from the surrounding mass fluctuations. This is
especially true at the edge of the field where noisy features in the
estimated map have been suppressed by the non-local solution provided
by the self-consistent method.  In particular, the features seen in
the lower left and lower right corners due to a large degree of
obscuration by the two brightest stars in the field (which is
difficult to properly account for) have been suppressed significantly.
In addition, the large dip seen in the top left corner caused by an
excessive number of background sources has been slightly reduced.

The error map calculated from the width of the probability
distribution in equation (\ref{eq_like_mag}) indicates that the
significance of the peak in the self-consistent map is $6\sigma$.  If
we include the extra uncertainty due to contamination by cluster
objects at the predicted level of 19\% as well as the error in
the unlensed surface number density, this significance drops to
$4\sigma$.  Note that the peaks in both plots appear lower than the
peak value implied by the radial analyses. This is a direct
consequence of the relatively low resolution of the $12\times 12$ grid
of bins which has averaged over the central cluster region and
effectively smoothed out the peak. This same reason describes the
apparent discrepancy with the shear-derived mass maps of TKD and B00,
both of which cover a much smaller field of view ($\sim 1.5' \times
1.5'$) and both of which imply a peak of $\kappa\simeq1.3$.

This resolution effect in the maps does not affect the shape of the radial
mass profiles because it occurs on a scale smaller than
the radial bin size. Section \ref{sec_radial_mass_profs} verifies
this since the profiles obtained agree with the power-law model
which predicts a central surface mass density of $\kappa\simeq1.4$.
Furthermore, the local estimator agrees with the non-local estimator
which would certainly not be the case if the bin resolution caused an
effective smoothing of the profile. The cumulative mass measurements
derived from the profiles are also therefore not affected.

The topology of the structure seen in the mass maps is very similar to
the distribution of cluster light (Figure \ref{clust_gal_light}) and
the cluster galaxy number density (Figure \ref{cluster_2d_no_dens}).
In the mass map, we find a relatively compact core with extensions
toward the north-west and south-west (toward the top-right and
bottom-right in Figure \ref{2d_kappa_R} respectively).  In comparing
these extensions with the light and number density distributions, it
is noticeable that the light traces the north-west extension more than
the south-west, whereas the number density traces the south-west
extension more than the north-west. The explanation is that in the
north-west, there are fewer galaxies, forming an almost distinct
sub-clump, but they are relatively bright.In contrast, the south-west
extension harbours a higher number of fainter cluster members.

Comparison of our mass map with the X-ray map of Soucail et al. (2000)
again shows very similar structure. The X-ray contours show an
elongation along the south-west direction while a separate peak of
X-ray emission is seen toward the north-west, coinciding with the
sub-clump of large cluster galaxies noted above.  Soucail et
al. suggest that the emission originates from one of two possible
sources. The closest to the centre of the emission is a large cluster
galaxy lying at a redshift of $z=0.4017$ while the other is a
foreground star-forming galaxy at $z=0.2132$. Given our detection of
mass and the concentration of large cluster galaxies in this region,
it seems plausible to explain at least part of the X-ray emission as
being due to cluster gas. The fact that the X-ray map shows a distinct
peak suggests that this is actually a separate mass clump rather than
an extension. The shear map of TKD shows no evidence of there being
such a sub-clump although B00's map suggests that extra mass lies in
this vicinity. Unfortunately, neither of these shear maps cover a wide
enough field of view for this to be properly verified.

\subsection{Radial mass profiles}
\label{sec_radial_mass_profs}

\begin{figure}
\vspace{4mm}
\epsfxsize=82mm
{\hfill
\epsfbox{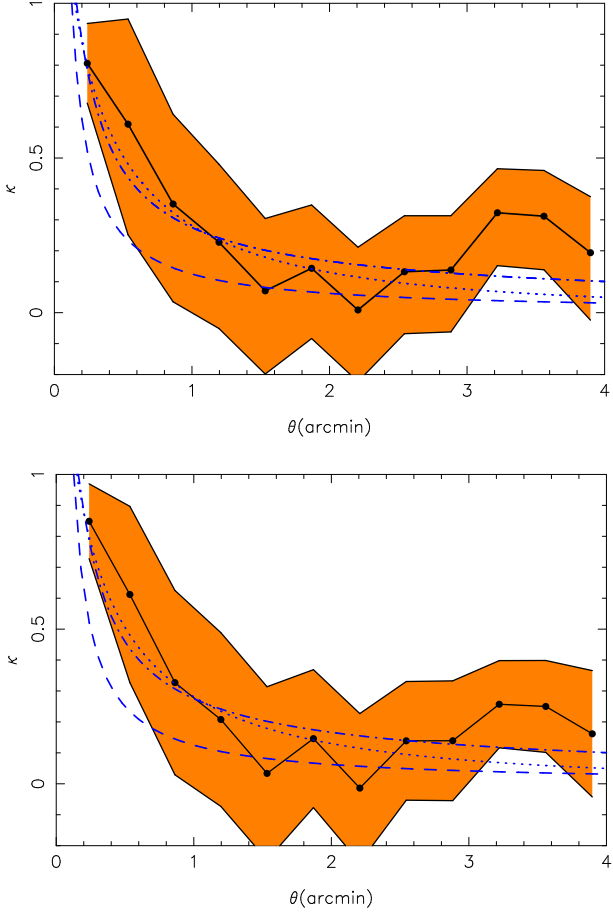}
\hfill}
\epsfverbosetrue
\caption{\small {\em Top}: Locally estimated radial mass profile for
$R$ band sources binned using annuli shown in Figure \ref{masks}. The
shaded region shows the $1\sigma$ uncertainty which allows for shot
noise, source clustering, uncertainty in background count
normalisation and 20\% cluster contamination. {\em Bottom}:
Axi-symmetric solution for the $R$ band.  Both plots show the
isothermal (dashes), power-law (dot-dashes) and NFW (dots) models
fitted to the first 9 data points (see Section
\ref{sec_R_band_fits}).}
\label{kap_R}
\end{figure}

In Figure \ref{kap_R} we show the radial mass profile calculated for
the $R$ band sources using both the local estimator and the
axi-symmetric method assuming the single number count slope model. For
calculation of the axi-symmetric profile, the value of the shear in
the first bin is set to $\gamma_1=0.25$ (see Section
\ref{sec_axi_symm_est}). The shaded regions in both plots show the
$1\sigma$ uncertainty accounting for shot noise, source clustering,
uncertainty in background count normalisation and 19\% cluster member
contamination.  Similar to the 2D mass distribution, we find that the
last 3 data points in the axi-symmetric solution are slightly
suppressed compared to the local estimator.

The fitted profiles of Section
\ref{sec_R_band_fits} plotted in Figure \ref{kap_R} again show that
the power-law and NFW models give a better fit to the results. Both profiles
exhibit an excess of mass at $3''<r<4''$. This is
attributed to the noisy features seen in the 2D mass plots at the
edge of the field, and just pushes the measured mass at this
radius to a value inconsistent with the NFW and power-law fit.

\subsection{Cumulative mass profile}
\label{sec_cum_mass}

To convert from $\kappa$ to real projected mass, we normalise to the
amount of mass contained in the disk traced by the observed
arcs. Since the enclosed mass can be calculated to a high accuracy,
normalising in this way provides a much more reliable scaling than
using a mean redshift estimated for the background population. Using
B00's measurement of the redshift of the arcs at $z=1.657$, the mass
within a radius of $r_{arc}=30''$ from the centre of the cluster can
be calculated to be $1.19\times 10^{14} h^{-1}{\rm M}_{\odot}$ for an
Einstein de Sitter (EdS) Universe. The cosmological dependence of this
result is weak to the extent that for an $\Omega=0.3$, $\Lambda=0.7$
cosmology, this enclosed mass increases by only 5\%. This scaling
applies generally hence all aperture masses quoted hereafter assume
an EdS cosmology.

\begin{figure}
\vspace{4mm}
\epsfxsize=82mm
{\hfill
\epsfbox{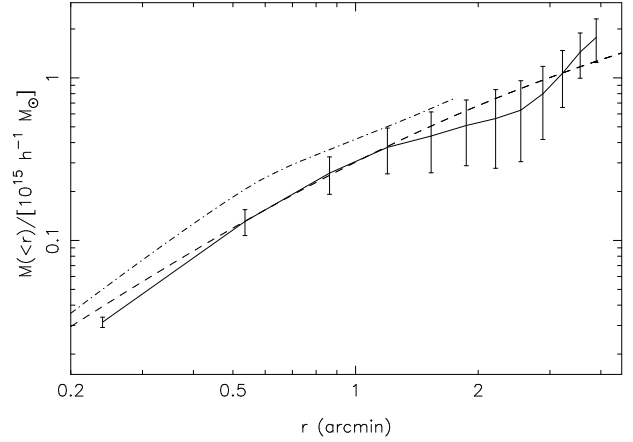}
\hfill}
\epsfverbosetrue
\caption{\small Cumulative projected mass calculated by normalising to
the amount of mass enclosed within the circle traced by the observed arcs.
Also plotted is the NFW model of Section \ref{sec_R_band_fits} (dashes)
and the $B$ band results of van Kampen (1998) (dot-dashes).}
\label{cum_mass}
\end{figure}

Scaling our radial $\kappa$ profile with this normalisation, we sum
the mass in each bin to produce the cumulative projected mass profile
in Figure \ref{cum_mass}. Out to a radius of $2.9'$ ($0.54 \mpcoh$ for
EdS), we measure a projected mass of $(8.1\pm 3.2)\times 10^{14}
h^{-1}{\rm M}_{\odot}$. The $1\sigma$ error here again includes shot
noise, source clustering, uncertainty in background count
normalisation and contamination from cluster members at the 19\%
level.  If we apply the dual slope model, we find a mass of $(8.2\pm
3.2)\times 10^{14} h^{-1}{\rm M}_{\odot}$, re-iterating the statement
made in Section \ref{sec_R_band_fits} that the lensing signal in the
$R$ band is dominated by galaxies in the $R<26$ magnitude range where
the slope $\beta_R=0.80$ applies.

Approximating the integrated NFW profile, the
projected mass within a radius $R$ scales as ${\rm M}(<R)=2.9\times
10^{14}(R/1')^{1.3-0.5\lg (R/1')}h^{-1}{\rm M}_{\odot}$.
With this scaling relation for the cumulative mass, we can readily
compare our results with other authors. Considering existing lensing
measurements for the moment, the measurement by B00 of the mass
contained within the arcs obviously agrees with our estimation by
default since this is essentially the mass we normalise to.  Using the
redshift of the arc from B00, TKD's estimate of the cluster mass
contained within a radius of $107 \kpcoh$ is $(1.553\pm0.002)\times10^{14}
h^{-1}{\rm M}_{\odot}$.  At this radius, our NFW estimate for the enclosed
mass is $(1.3\pm0.3)\times10^{14} h^{-1}{\rm M}_{\odot}$. Extending
to slightly larger radii, the magnification analysis of van Kampen
(1998) finds a projected mass of $(6.7\pm2.0)\times10^{14} h^{-1}{\rm
M}_{\odot}$ within $300\kpcoh$ compared to our NFW mass of
$(5.0\pm1.7)\times10^{14} h^{-1}{\rm M}_{\odot}$.  On even larger scales, the
weak shear study of Bonnet, Mellier\& Fort (1994) measures a mass of
$(1.6\pm0.4)\times10^{15} h^{-1}{\rm M}_{\odot}$ within $R=1.5\mpcoh$ compared
with our prediction of $(1.7\pm0.7)\times10^{15} h^{-1}{\rm M}_{\odot}$.

In terms of mass-to-light, using the surface flux density of cluster
galaxies plotted in Figure \ref{clust_gal_light}, we find a ratio of
M/L$_B=330\pm30$ (total mass/total light in $B$) within a radius of
$0.5'$. This error accounts for uncertainty in the photometric zero
point but does not reflect the fact that this is effectively an upper
limit as a result of potentially missing cluster galaxies in our
selection process (Section \ref{sec_obsel}). Again, since our mass
within this radius agrees with that of B00, we find an almost
identical result to B00's result of M/L$_B=320\pm30$.  On a slightly
larger scale, we measure M/L$_B=480\pm180$ within $1'$ where now the
error includes additional uncertainty from the mass error. Finally,
out to $2.9'$, we find M/L$_B=470\pm190$. These latter two
measurements are consistent with a mass-to-light ratio which increases
on larger scales although a scale-independent ratio, as found by
Kochanski, Dell' Antonio \& Tyson (1996), remains plausible given the
errors.

In Figure \ref{cum_mass}, we also plot the only other magnification
based cumulative mass profile of CL0024$+$1654 which exists to date;
that of van Kampen (1998). This was derived from the $B$ band
observations of the cluster by Fort, Mellier \& Dantel-Fort (1997).
We choose the $B$ band data in favour of the $I$ which suffered
contamination from an unusual speckle beam pattern coming from stray
light in the telescope optics (Fort 2001, private communication).
This resulted in an over-exaggerated zone of depletion in the $I$ band
radial number counts.  Notice that van Kampen's $B$ band mass profile
agrees very well with our data points in terms of its shape but that
its normalisation is higher by $\sim 50\%$.

Turning to alternative methods for the determination of cluster mass,
Czoske et al. (1999) using galaxy dynamics from 227 spectra measure
the 3D mass of CL0024$+$1654 to be $(1.4\pm0.3)\times10^{14}
h^{-1}{\rm M}_{\odot}$ within a radius of $500\kpcoh$. Converting this
to a projected mass using the isothermal sphere model they assume,
yields a value of $(2.2\pm0.5)\times10^{14} h^{-1}{\rm
M}_{\odot}$. This agrees with the X-ray temperature measurements by
Soucail et al. (2000) who measure a projected mass of
$2.3^{+2.0}_{-0.8}\times10^{14} h^{-1}{\rm M}_{\odot}$ within the same
radius. Our scaling relation gives a projected mass enclosed by this
radius of $(8.0\pm3.1)\times10^{14} h^{-1}{\rm M}_{\odot}$.  It should
be noted that the later redshift measurements of Czoske et al. (2001a)
indicate that CL0024$+$1654 is a perturbed system. This implies that
mass measurements of the cluster which assume dynamical equilibrium
are probably not reliable. We discuss the significance of this in
Section \ref{sec_summary}.

\section{The $U$ band galaxy population}
\label{sec_U_band_pop}

The role of the $U$ band population thus far has been to assist in the
selection of $R$ band background galaxies through colour cuts.  In
this section, we test the feasibility of using the $U$ band
for detection of lens magnification through number counts.  We
wish to re-examine the claim made by R01 that the number count slope
of the background field galaxy sample becomes flatter than the lens
invariant slope, $\beta=1$, at faint magnitudes.

\subsection{The $U$ band number count slope}
\label{sec_Uband_slopes}

Current evidence regarding the notion of a break in the $U$ band field
galaxy number count slope at faint magnitudes is tenuous. At
magnitudes brighter than $U_{\rm AB}\simeq 25.5$, there is good
convergence. All studies agree that at these brighter magnitudes, the
number count slope is relatively steep.  Pozzetti et al. (1998), in
analysing the Hubble Deep Field North (HDF-North), find that for
$23<U_{\rm AB}<25.5$, $\beta_U=1.00$ whereas Hogg et al. (1997)
conclude that $\beta_U=1.17$ for $U_{\rm AB}<25.5$ from ground based
observations.

At fainter magnitudes, Pozzetti et al. (1998) measure a much flatter
slope of $\beta_U=0.34$ at $U_{\rm AB}=25.8$ from the HDF-North.
Similarly, at the same magnitude, Metcalfe et al. (2001), using a
combination of the ground based William Herschel Deep Field and the
HDF-North and HDF-South, report a slope of $\beta_U=0.38$.  This
appears to contrast with the results of Volonteri et al. (2000) who
measure a slope of $\beta_U=1$ up to $U_{\rm AB}\simeq 26.5$ with the
HDF-South.

The depth of our observations in the $U$ band extend to approximately
the suggested $U$ band break magnitude, which at first sight, appears
to extinguish any hopes of pushing the search beyond current limits.
However, we are in a fortuitous position for two reasons. The first is
that our analysis merely rests upon the {\em relative} depletion of
galaxy numbers. As Gray et al. (2000) demonstrate, this means that
provided the completeness characteristics of the unlensed reference
number counts are the same as the lensed field, the completeness
function drops out of equation (\ref{eq_mag_eqn}).  Our reference
counts are taken from the edges of our cluster field so by definition
have the same completeness characteristics.  In the same way that the
completeness function drops out of equation (\ref{eq_mag_eqn}), it
also drops out of equations (\ref{eq_mag_eqn_dual_simple}) and
(\ref{eq_breaktest}). This is an important result and means that our
break search is not affected by incompleteness. We therefore need not
concern ourselves with completeness corrections such as those applied
to existing traditional number count measurements which may cause a
potentially large uncertainty at faint magnitudes.

Our second fortuity lies in the fact that our observations are of a
cluster-lensed field, enabling us to exploit the effect of flux
magnification on our background sample.  Taking the NFW model fit from
Section \ref{sec_R_band_fits}, this predicts that within a disk of
diameter $0.8'$ centred on the cluster, lens magnification by
CL0024$+$1654 pushes the effective limiting magnitude of our
observations by $\Delta m \geq 2.5$. Within a disk of diameter $2'$,
this becomes $\Delta m \geq 0.8$. We are therefore able to use our
dataset to search for a break in the $U$ band number counts some way
beyond the physical limiting magnitude of our data.

Note that this does not contradict the discussion earlier that
allowing for a dual slope in the $R$ band lensing analysis causes
little difference to the results obtained. It is true that both the
$U$ and $R$ band samples extend to approximately where the break
magnitude is thought to lie and it is also true that lens
magnification acts on both samples equally to increase the effective
depth of each.  In the case of the $R$ band sample however, the
determination of mass depends on the {\em joint} contribution from
both the steep and shallow number count slopes. In the search for the
$U$ band break as we discuss in detail below, these contributions are
considered {\em separately} so that the more numerous galaxies in the
steep slope region do not dominate those beyond the break
magnitude. In addition to this effect, the expected $U$ band change of
slope of $\beta_U\simeq 1 \rightarrow 0.3$ is more severe than the
expected $R$ band change of $\beta_U\simeq 0.8 \rightarrow 0.5$. This
makes detection of any faint slope depletion easier and hence further
increases the confidence limits we can place on our search results.

\subsection{Testing for a $U$ band break}
\label{sec_U_break}

We use two number count models to search for a break magnitude. The
first is a modification of the single slope number count scenario to
allow the slope to smoothly flatten off beyond the break magnitude. The
second uses the dual slope model of Section \ref{sec_dual_slope_mag},
allowing the faint slope to vary.

\subsubsection{Model 1}

The first model generalises the single slope number count model to include
an arbitrary cut-off at $U_0$;
\be
\label{eq_breaktest}
n'_U = n_{U} \mu^{\beta_U-1} \left[ \frac{1+10^{0.4 \beta_U \Delta U}}
	{1+\mu^{\beta_U} 10^{0.4 \beta_U \Delta U}}\right],
\ee
where $\Delta U= U_{lim}-U_0$ and $U_{lim}$ is the limiting magnitude
of the observations, $U_{lim}=25.7$. Well below the break scale, $U_0
\ll U_{lim}$, or for weak lensing $\mu \approx 1$, this reduces to the
usual scaling for lensing of a single slope number count distribution
in equation (\ref{eq_mag_eqn}). Above the break scale, equation
(\ref{eq_breaktest}) tends toward the scaling $n'=n \mu^{-1}$, for a
completely flat number count distribution. 

We take the best fit NFW profile determined from the $R$ band data to
provide the magnification in equation (\ref{eq_breaktest}).  The free
parameters of the model are therefore the break scale, $U_0$, and
$\beta_U$ which we fit by maximising the following likelihood function:
\be
\label{eq_chi_sq_fn}
{\cal L} \propto \left<\prod_i \exp\left(\frac{-[n'_{U,i} 
- p_{U,i}(U_0,\beta_U)-c_i]^2}{n_{U,i}}\right)\right>_c.
\ee
Here, $n'_{U,i}$ is the number of $U$ band objects observed in bin
$i$, $n_{U,i}$ is the number expected in the absence of lensing and
$p_{U,i}$ is the number predicted by the NFW profile fitted to the $R$
band data. The product acts over all annuli. Uncertainty due to
cluster and foreground galaxy contamination is incorporated through
the quantity $c_i$. The foreground galaxy component of $c_i$ is set at
3\% of $n_{U,i}$ for each bin, in accordance with the contamination
fraction discussed in Section \ref{sec_fgnd_contam}. The cluster
component of $c_i$ varies with choice of CLF. We therefore average
over different realisations of the CLF, weighting by the probability
distribution obtained from fitting to our known $U$ band cluster
counts (see Section \ref{sec_cluster_contam}). As in Section
\ref{sec_results}, the radial variation of cluster number count
density is taken to be $k/r$ where $k$ is set by normalising to the
contamination determined for each CLF realisation.

\begin{figure}
\epsfxsize=82mm
{\hfill
\epsfbox{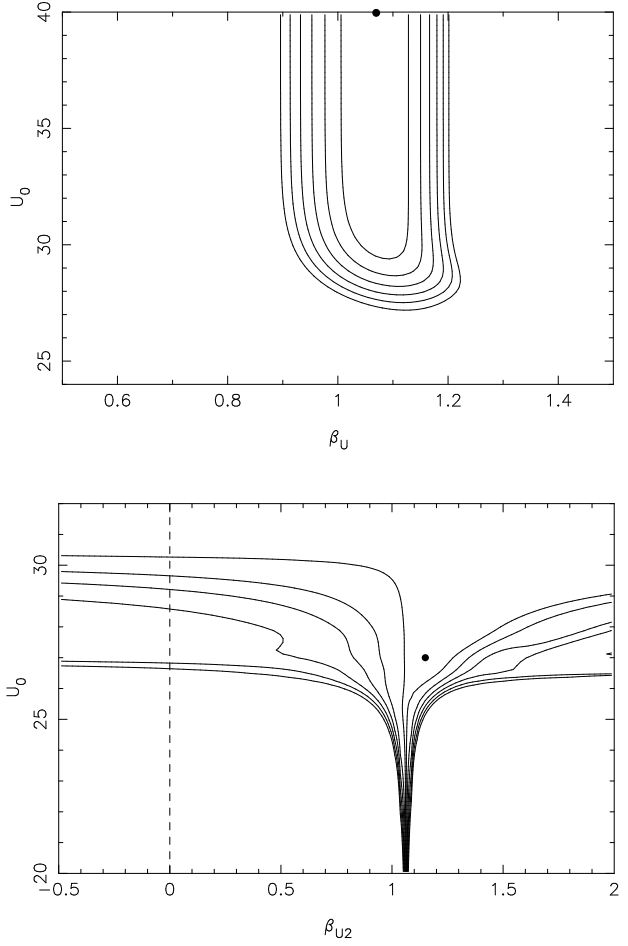}
\hfill}
\epsfverbosetrue
\caption{\small {\em Top}: $\chi^2$ distribution of the $U$ band number
count slope $\beta_U$ and the break scale $U_0$ assuming the number counts
flatten off completely at faint magnitudes. {\em Bottom}:  
$\chi^2$ distribution of the faint number count slope $\beta_{U2}$
and the break scale holding the bright slope at $\beta_{U1}=1.07$.
All contours are separated by $\Delta \chi^2 = 1$.}
\label{Ubreak}
\end{figure}

The top half of Figure \ref{Ubreak} shows the $\chi^2$ contours
obtained from the likelihood distribution of equation
(\ref{eq_chi_sq_fn}). With a 95\% confidence, our data rules out a
complete flattening of the $U$ band counts with a break magnitude
brighter than $U_{\rm AB}=27.3$. Furthermore, we measure the $U$ band
slope as $\beta_U=1.07\pm0.06$.  This is very close to the lensing
invariant slope $\beta=1$ which causes the increase in surface number
density due to flux magnification to exactly cancel the dilution
caused by magnification of their inter-spacing. The claim made in
Section \ref{sec_no_count_fitting} that the $U$ band population is
unsuitable for the detection of magnification from number counts is
thus substantiated.

\begin{figure}
\vspace{4mm}
\epsfxsize=82mm
{\hfill
\epsfbox{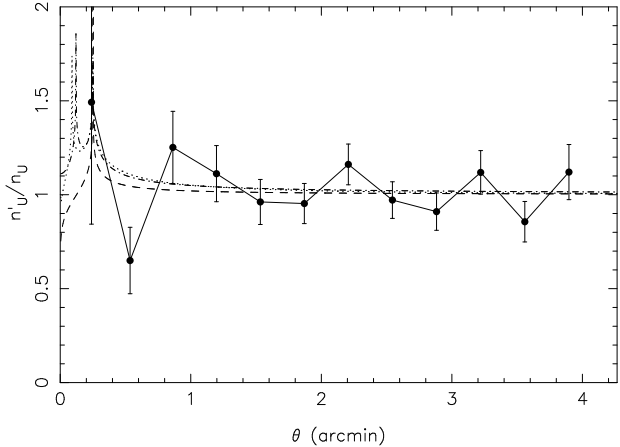}
\hfill}
\epsfverbosetrue
\caption{\small Radial number count profile in the $U$ band.
Superimposed are the best fit isothermal (dashed line), power-law
(dot-dashed line) and NFW (dotted line) models using the number count slope
$\beta_U=1.07$.  Error bars account for the error in $n_U$ and shot
noise.}
\label{n_n0_U}
\end{figure}

Figure \ref{n_n0_U} shows the $U$ band radial number count profile
observed.  In this plot, the model fits
determined in Section \ref{sec_R_band_fits} are used to predict counts
assuming $\beta_U=1.07$. The predictions and measurements are
clearly consistent with each other. There is little distinction between
any of the predicted profiles due to the near lensing invariant
slope.  This serves to demonstrate the
insignificant depletion signal imposed on the number counts in the
$U$ band.

\subsubsection{Model 2}

The second method we use to search for a break assumes the dual number
count slope model described in Section \ref{sec_dual_slope_mag}.  For
the purpose of our analysis in this section, we hold the bright slope
at the value determined previously, $\beta_{U1}=1.07$. The faint slope
$\beta_{U2}$ and the break magnitude are allowed to vary in our
minimisation.

The lower half of Figure \ref{Ubreak} shows the results of this
minimisation. Slightly more relaxed than the previous result, the data
rule out a complete flattening ($\beta_{U2}=0$) brighter than $U_{\rm
AB}=26.6$ at a confidence level of 95\% . As the faint slope is
allowed to steepen, this limit slips to brighter magnitudes as a
result of the degeneracy between both parameters (Pozzetti at
al. 1998).  However, a slope of $\beta_{U2}=0.4$, the value reported
by Pozzetti et al. (1998) and Metcalfe et
al. (2001) as being applicable fainter than $U_{\rm
AB}\simeq25.5\rightarrow 26$, can be ruled out to $U_{\rm AB}<26.4$
with 95\% confidence. Pushing this limit further still, the contours
in Figure \ref{Ubreak} show that a slope even as steep as
$\beta_{U2}=0.8$ can be ruled out with 95\% confidence to $U_{\rm
AB}<26.0$. Finally, at $\beta_{U2}=\beta_{U1}$, where the degeneracy
between the faint and steep slope is 100\%, the break magnitude cannot
be constrained at all.

We note that these results are also inconsistent with the findings of
R01 who suggest a {\em combined} slope (ie. the effective single slope
causing the same level of depletion as a dual-slope model,
steeper than the faint end slope) of $\beta\simeq 0.5$.

\section{Summary and Discussion}
\label{sec_summary}

From $U$ and $R$ band observations of the cluster CL0024$+$1654, we
have selected a background sample of galaxies in both bands using
colour information. The cluster member contaminants identified by
matching with the Czoske et al. (2001a) redshift survey of the field
have been discarded from the $U$ band. We have computed the cluster
luminosity function of CL0024$+$1654 in both bands. This has been used
in conjunction with the $U$ and $R$ field galaxy luminosity function
from the CNOC2 survey (Lin et al. 1999) to estimate foreground and
cluster contamination of our background samples. In the $R$ band, we
estimate a total contamination of $21\%$ in contrast to a smaller
$12\%$ in the $U$, where the cluster component is calculated from
our own fitted CLF.

The shallower number count slope observed in the $R$ band sample makes
this suitable for an investigation into the lens magnification induced
depletion of background number counts. Depletion in this sample has
indeed been detected and used to measure the radial and two
dimensional distribution of mass in the cluster.  Out to a radius of
$0.54\mpcoh$, we measure a total projected mass of $(8.1\pm 3.2)\times
10^{14} h^{-1}{\rm M}_{\odot}$ where the $1\sigma$ error includes shot
noise, source clustering, uncertainty in background count
normalisation and contamination from cluster and foreground
galaxies. We find that this result increases by merely $\sim 2\%$ when
we allow for a flattening in the $R$ band number count slope at
$R_{\rm AB}=26$. Claims of such a change in slope have been reported
by several authors (eg. Pozzetti et al. 1998; Metcalfe et al. 2001),
however allowing for it in our depletion analysis makes little
difference due to the lensing signal being dominated by galaxies
brighter than the break.

Converting the $R$ band flux from selected cluster galaxies to the $B$
band, we find a mass-to-light ratio of M/L$_B=330\pm30$ inside an
aperture of radius $0.5'$ centred on the cluster.  Since our selection
process will have inevitably missed cluster galaxies, neglecting their
contribution to the total luminosity, this is
effectively an upper limit. On a slightly larger scale, we measure
M/L$_B=480\pm180$ within $1'$ and also M/L$_B=470\pm190$ within $2.9'$
($0.54\mpcoh$). These latter two measurements are consistent with a
mass-to-light ratio which increases on larger scales although a
scale-independent ratio remains plausible given the errors.

We have compared the observed radial depletion with that expected from
an isothermal, power-law and NFW mass profile by fitting the predicted
magnification of each. The NFW model provides the best fit to our data
however there is little distinction between this and the power-law
model. This agrees with the findings of Tyson, Kochanski \& Dell'
Antonio (1998) as does the mass of $1.3\times 10^{14}h^{-1}{\rm
M}_{\odot}$ we measure within the disk described by the large arcs
compared to their estimate of $(1.553\pm0.002)\times 10^{14}h^{-1}{\rm
M}_{\odot}$.  Approximating the fitted cumulative NFW model, the
projected mass contained within a radius $R$ scales approximately as
${\rm M}(<R)=2.9\times 10^{14}(R/1')^{1.3-0.5\lg (R/1')}h^{-1}{\rm
M}_{\odot}$. Together with the cluster galaxy light and number density
distribution, our 2D mass maps suggest the existence of a separate
sub-clump of mass just north-west of the cluster centre. This claim is
strengthened by the detection of distinct X-ray emission in this area
(Soucail et al. 2000) as well as the suggestion of extra mass from the
shear mass map of Broadhurst et al. (2000).

In fitting the number counts to the isothermal sphere mass model, we
find that the fitted critical radius is smaller (although not
significantly given the error) than the radius of the circle traced by
the observed large arcs. This might be due to the fact that our
selected background population of sources lies at a lower mean
redshift than the lensed galaxy forming the arc at $z=1.675$
(Broadhurst et al. 2000). The mean background source redshift in an
80\% confidence interval is inferred to be within
$0.40<z_{mean}<1.19$. The depletion analysis of CL0024$+$1654 by Fort,
Mellier \& Dantel-Fort (1997) in the $B$ band finds a relatively wide
depletion profile (see Section \ref{sec_cum_mass} regarding
contamination of their $I$ band data).  Simultaneously fitting to a
maximum of five isothermal sphere models, they find that a combination
of various background galaxy populations is required to explain the
wide zone of depletion. In $B$, this results in the mean background
source redshifts ranging over $z=0.9^{+0.1}_{-0.1}$ to
$z=3.0^{+1.8}_{-0.5}$ with 42\% of sources lying at $z=0.9$.  This is
a little surprising given that we find an adequate fit to our observed
depletion profile assuming the existence of only one critical
line. The discrepancy would perhaps be at least partially resolved by
fitting the Fort, Mellier \& Dantel-Fort data to an NFW profile since
this accommodates a larger zone of depletion compared to the
isothermal sphere model.

Comparison of our results or indeed any of the existing lensing
results with those measured using cluster galaxy dynamics or X-ray
temperatures shows a large discrepancy. We predict approximately $3.5$
times as much projected mass as the X-ray and dynamical measurements
imply.  Soucail et al. (2000) discuss that one explanation for this
discrepancy may come from the inability to correctly measure lens
shear combined with a lack of accurate knowledge of the background
source redshift distribution.  While this bears some truth in general,
it is not the case here.  The determination of the arc redshift by
Broadhurst et al. (2000) makes the measurement of projected mass
contained within the arc radius a robust one; this is a result which
does not depend on the detection of shear through weak lensing and 
also knowing the redshift of the arced galaxy breaks the source
redshift degeneracy.  The projected mass contained within the arc
radius according to the X-ray measurements of Soucail et al. is
$(0.5\pm 0.3)\times 10^{14} h^{-1}{\rm M}_{\odot}$. This is still a
factor of nearly three times smaller than the arc predicted mass.

An alternative scenario which provides a more satisfying answer to the
evidence gained thus far is alignment of additional foreground and/or
background mass along the line of sight to CL0024$+$1654. The Czoske
et al. (2001a) redshift survey identifies a group of galaxies lying
just in front of the cluster as well as a pair of groups lying behind
it. During the course of preparing this paper, new evidence emerged
regarding the perturbed galaxy dynamics of CL0024$+$1654.  Comparing
numerical simulations with their redshift measurements, Czoske et
al. (2001b) conclude that the cluster could in fact be the result of a
high speed collision of two smaller clusters along the line of
sight. This would certainly explain why lensing observes a factor of
three times as much mass. It also explains why early measurements of
velocity dispersion were very large (see Dressler \& Gunn 1992 and
references therein); these were based on $\sim 10$ times fewer
redshifts and hence failed to properly resolve the cluster's true
dynamical state.

Finally, the $U$ band selected background sample indicates a
near-lensing-invariant slope of $\beta=1.07\pm0.06$ and hence does not
exhibit any noticeable sign of depletion. We have used the fact that
lens magnification of our $U$ band sample allows us to search deeper
than its physical limiting magnitudes thus facilitating the search for
a change of slope in the $U$ band number counts at faint
magnitudes. Up to $U_{\rm AB}\leq 26.6$, we can rule out a complete
flattening to 95\% confidence. Furthermore, we can rule out the
existence of a change of slope of $\beta=1\rightarrow 0.4$ reported by
Pozzetti et al. (1998) and Metcalfe et al. (2001) with a confidence of
95\% up to $U_{\rm AB}\leq26.4$.

Our findings also contradict the results of R01 who claim to have
measured $U$ band depletion and hence slope flattening.  Although we
started with the same observations as R01, we applied more stringent
criteria to the object extraction and carried out a thorough $U$ band
break analysis which allowed for contamination by cluster and foreground
galaxies. It is intriguing that the HDF-South $U$ band number counts
reported by Volonteri et al. (2000) also support our result. They too
find no evidence of flattening in the $U$ band counts up to $U_{\rm
AB}simeq26.5$. This will most likely remain an unresolved issue until
deeper observations, particularly in a cluster environment where lens
magnification provides natural assistance, are obtained.

\bigskip
\noindent{\bf ACKNOWLEDGEMENTS} \bib\strut

\noindent
SD thanks PPARC for financial support as a PDRA. ANT is a PPARC
Advanced Fellow.  Thanks to Chris Pearson for helpful input from his
source number count models. Finally, we thank Ian Smail as referee of
this paper for his constructive comments. This work was supported in
part by The Icelandic Research Council and The Research Fund of the
University of Iceland.  The Nordic Optical Telescope is operated on
the island of La Palma jointly by Denmark, Finland, Iceland, Norway,
and Sweden, in the Spanish Observatorio del Roque de los Muchachos of
the Instituto de Astrofisica de Canarias.

\bigskip
\noindent{\bf REFERENCES}
\bib \strut

\bib Bartelmann M., 1996, A\&A, 313, 697

\bib Beijersbergen M., Hoekstra H., van Dokkum P.G., van der Hulst T.,
	2001, submitted to MNRAS, astro-ph/0106354

\bib Bertin E. \& Arnouts S., 1996, A\&AS, 117, 393

\bib Blandford R.D. \& Narayan R., 1992, ARA\&A, 20, 311

\bib Bonnet H., Mellier Y. \& Fort B., 1994, ApJ, 427, L83

\bib Broadhurst T.J., Taylor A.N. \& Peacock J.A., 1995, ApJ, 438, 49 (BTP)

\bib Broadhurst T.J., Huang X., Frye B. \& Ellis R., 2000, ApJ, 534, L15 (B00)

\bib Cohen J., Hogg D., Blandford R., Cowie L., Hu E., Songalia A.,
	Shopbell P., Richberg K., 2000, ApJ, 538, 29

\bib Coles P. \& Jones B., 1991, MNRAS, 248, 1

\bib Czoske O., Soucail G., Kneib J.-P., Bridges T., Cuillandre J.-C.,
	Mellier Y., 1999, "Gravitational Lensing: Recent Progress and 
	Future Goals", Boston University, July 1999, ed. T. G. 
	Brainerd and C. S. Kochanek

\bib Czoske O., Kneib J.-P., Soucail G., Bridges T., Mellier Y., 
	Cuillandre J.-C., 2001a, A\&A, 372, 391

\bib Czoske O., Moore B., Kneib J.-P., Soucail G., 2001b, submitted to
	A\&A, astro-ph/0111118

\bib Dressler A. \& Gunn J.E., 1992, ApJS, 78, 1

\bib Dressler A., Oemler A., Couch W.J., Smail I., Ellis R.S.,
	Barger A., Butcher H., Poggianti B.M., Sharples R.M., 
	1997, ApJ, 490, 577

\bib Dressler A., Schneider D.P. \& Gunn J.E., 1985, ApJ, 294, 70

\bib Driver S.P., Phillips S., Davies J.I., Morgan I., Disney M.J., 1994,
	MNRAS, 268, 393

\bib Dye S. \& Taylor A.N., 1998, MNRAS, 300, L23

\bib Fort B., Mellier Y. \& Dantel-Fort M., 1997, A\&A, 321, 353

\bib Fukugita M., Shimasaku K. \& Ichikawa T., 1995, PASP, 107, 945

\bib Garilli B., Maccagni D. \& Andreon S., 1999, A\&A, 342, 408

\bib Ghigna S., Moore B., Governato F., Lake G., Quinn T., Stadel J., 
	1998, MNRAS, 300, 146

\bib Gray M.E., Ellis R.S., Refregier A., B\'{e}zecourt J., 
 	McMahon R.G., Beckett M.G., Mackay C.D., Hoenig M.D.,
	2000, MNRAS, 318, 573

\bib Hogg D.W., Pahre M.A., McCarthy J.K., Cohen J.G., Blandford R.,
	Smail I., Soifer B.T., 1997, MNRAS, 288, 404

\bib Hudon J.D. \& Lilly S.J., 1996, ApJ, 469, 519

\bib Kaiser N., Squires G. \& Broadhurst T., 1995, ApJ, 449, 460

\bib Kassiola A., Kovner I. \& Fort B., 1992, ApJ, 400, 41

\bib Kinney A.L., Calzetti D., Bohlin R.C., McQuade K., Storchi-Bergmann T.,
	Schmitt H.R., 1996, ApJ, 467, 38

\bib Kochanski G.P., Dell' Antonio I.P. \& Tyson J.A., 1996,
	AAS Meeting, 189, 73.02

\bib Koo D.C., 1988, in Large Scale Motions in the Universe, eds.
	V.C. Rubin \& G.V. Coyne, Princeton University Press, 513

\bib Kovner I., 1990, in Gravitational Lensing, eds. Y. Mellier,
	B. Fort \& G. Soucail, Springer, 16

\bib Lin H., Yee H.K.C., Carlberg R.G., Morris S.L., Sawicki M.,
	Patton D.R., Wirth G., Shepherd C.W., 1999, ApJ, 518, 533

\bib Metcalfe N., Shanks T., Campos A., McCracken H.J., Fong R., 2001,
	MNRAS, 323, 795

\bib Navarro J.F., Frenk C.S. \& White S.D.M., 1997, 
	ApJ, 490, 493 (NFW)

\bib Paolillo M., Andreon S., Longo G., Puddu E., Gal R.R., 
	Scaramella R., Djorgovski S.G., de Carvalho R., 
	2001, A\&A, 367, 59

\bib Piranomonte S., Paolillo M., Andreon S., Longo G., Puddu E.,
	Scaramella R., Gal R., Djorgovski S.G., 2000, IAP Conference
	``Constructing the Universe with Clusters of Galaxies'', 
	Paris, July 2000

\bib Pozzetti L., Madau P., Zamorani G., Ferguson H.C., Bruzual A.G.,
	1998, MNRAS, 298, 1133

\bib R\"{o}gnvaldsson \"{O}.E. et al., 2001, MNRAS, 322, 131 (R01)

\bib Sandage A., Tammann G.A., Yahil A., 1979, ApJ, 232, 352

\bib Schechter P., 1976, ApJ, 203, 297

\bib Schlegel D., Finkbeiner D. \& Davis M., 1998, ApJ, 500, 525

\bib Schneider P., Ehlers J., Falco E.E., 1993, Gravitational Lenses,
	New York: Springer

\bib Smail I., Hogg D., Yan L., Cohen J., 1995, ApJ, 449, L105

\bib Smail I., Dressler A., Couch W.J., Ellis R.S., Oemler A., 
	Butcher H., Sharples R.M., 1997, ApJ, 110, 213

\bib Soucail G., Ot, N., B\"{o}hringer H., Czoske O., Hattori M., 
	Mellier Y., 2000, A\&A, 355, 433

\bib Taylor A.N., Dye S., Broadhurst T.J., Benitez N., van Kampen E.,
	1998, ApJ, 501, 539 (T98)

\bib Tyson J.A., Kochanski G.P. \& Dell' Antonio I.P., 1998,
	ApJ, 498, L107 (TKD)

\bib van Kampen E., 1998, MNRAS, 301, 389

\bib Volonteri M., Saracco P., Chincarini G., Bolzonella M., 2000,
	A\&A, 362, 487

\bib Wallington S., Kochanek C.S. \& Koo D.C., 1995, ApJ, 441, 58

\bib Williams et al., 1996, AJ, 112, 1335

\bib Wilson G., Smail I., Ellis R.S., Couch W.J., 1997, MNRAS, 284, 915

\end{document}